\documentclass[11pt,a4paper]{article}                           

\usepackage{amsmath}
\usepackage{amssymb}
\usepackage[usenames]{color} 
\usepackage{multirow}
\usepackage{graphicx}
\usepackage{epstopdf}
\usepackage{array}
\usepackage{hhline}
\usepackage{cite}
\usepackage{microtype,colortbl}
\usepackage[bf]{caption}
\usepackage{color}
\usepackage[usenames,dvipsnames]{xcolor}
\usepackage{subcaption}
\usepackage{pstool}
\usepackage{tikz}
\usepackage{ytableau}
\usepackage{latexsym,stmaryrd}
\usepackage[bookmarks=false]{hyperref}
 
\usepackage{ifpdf}
\def\hybrid{\topmargin -20pt    \oddsidemargin 0pt
        \headheight 0pt \headsep 0pt
        \textwidth 6.25in       
        \textheight 9 in       
        \marginparwidth .875in
        \parskip 5pt plus 1pt 
          \jot = 1.5ex
   }
\hybrid
\numberwithin{equation}{section}
\numberwithin{table}{section}

\newcommand{\beq}{\begin{equation}}  \newcommand{\eeq}{\end{equation}}
\newcommand{\bal}{\begin{aligned}}   \newcommand{\eal}{\end{aligned}}
\newcommand{\bea}{\begin{eqnarray}}  \newcommand{\eea}{\end{eqnarray}}

\newcommand{\bmat}{\left(\begin{array}}
\newcommand{\emat}{\end{array}\right)}

\newcommand{\bbC}{\mathbb{C}}
\newcommand{\bbR}{\mathbb{R}}

\newcommand{\nn}{\nonumber}

\newcommand{\cE}{\mathcal{E}}

\newcommand{\cL}{\mathcal{L}}

\newcommand{\cN}{\mathcal{N}}

\newcommand{\cI}{\mathcal{I}}
\newcommand{\cJ}{\mathcal{J}}

\newcommand{\cM}{\mathcal M}

\renewcommand{\Im}{\mathrm{Im}\,}
\renewcommand{\Re}{\mathrm{Re}\,}

\newcommand{\I}{\text{Im}}
\newcommand{\R}{\text{Re}}

\usepackage{cancel}

\newcommand{\be}{\begin{equation}}
\newcommand{\ee}{\end{equation}}

\newcommand{\spanC}[1]{\mathrm{span}_\bbC\{#1\}}

\definecolor{Gray}{gray}{0.95}

\begin{document}

\vspace*{-1.5cm}
\begin{flushright}    
  {
  }
\end{flushright}

\vspace{3cm}
\begin{center}        
  {\LARGE Infinite Distances and the Axion Weak Gravity Conjecture}
\end{center}

\vspace{0.5cm}
\begin{center}        
{\large  Thomas W.~Grimm and Damian van de Heisteeg}
\end{center}

\vspace{0.15cm}
\begin{center}        
\emph{ Institute for Theoretical Physics \\
Utrecht University\\ 
Princetonplein 5, 3584 CE Utrecht, The Netherlands}\\[.3cm]

\end{center}

\vspace{3cm}


\begin{abstract}
\noindent
The axion Weak Gravity Conjecture implies that when parametrically increasing the 
axion decay constants, instanton corrections become increasingly important. 
We provide strong evidence for the validity of this conjecture by studying 
the couplings of R-R axions arising in Calabi-Yau compactifications of Type IIA string theory. 
Specifically, we consider all possible infinite distance limits in 
complex structure moduli space and identify the axion decay constants that grow 
parametrically in a certain path-independent way.
We then argue that for each of these limits a tower of D2-brane instantons with decreasing 
actions can be identified. These instantons ensure that the convex hull condition relevant for the multi-axion Weak Gravity Conjecture 
cannot be violated parametrically.  
To argue for the existence of such instantons we employ and generalize recent insights about the 
Swampland Distance Conjecture. Our results are general and not restricted to specific examples, since we 
use general results about the growth of the Hodge metric 
and the sl(2)-splittings of the three-form cohomology associated to each limit.

\end{abstract}

\thispagestyle{empty}
\clearpage

\setcounter{page}{1}


\newpage

\begingroup
  \flushbottom
  \tableofcontents
\endgroup
\newpage

\section{Introduction}
\label{sec:intro}

A generic feature of four-dimensional string compactifications is the existence  
of scalar fields with approximate shift symmetries appearing in the effective action. 
Such axions appear in many phenomenologically motivated models, including models of 
cosmic inflation, in which the axions are considered as inflaton fields \cite{Baumann:2014nda,Marsh:2015xka}. 
The axion decay constants $f$, determined by the 
kinetic term of the axions, are of crucial importance and, depending on the precise application,
required to have a certain range of values. For example, in large 
field inflationary models large values for $f$, comparable to the Planck mass, are often 
required. It has been suggested long ago, starting with \cite{Banks:2003sx}, that large axion 
decay constants cannot be obtained in string theory, while at the same time controlling the 
instanton corrections that need to be included in a consistent formulation of the effective theory. 
This observation is formalized in the axion Weak Gravity Conjecture \cite{ArkaniHamed:2006dz}, which states 
that the axion decay constant times the action of an appropriate instanton coupling to the axion is smaller 
than the Planck mass times the instanton charge. Clearly, it is notoriously hard to establish such a conjecture, since it requires 
to control the behaviour of the axion decay constants and then reliably determine the non-perturbative 
corrections to the effective action as recently reviewed in \cite{Palti:2019pca}. 

Recently, the Swampland Distance Conjecture \cite{Ooguri:2006in} has received much attention \cite{Baume:2016psm,Klaewer:2016kiy,Valenzuela:2016yny,Blumenhagen:2017cxt,Palti:2017elp,Hebecker:2017lxm,Lust:2017wrl,Cicoli:2018tcq,Grimm:2018ohb,Heidenreich:2018kpg,Blumenhagen:2018nts,Landete:2018kqf,Blumenhagen:2018hsh,Lee:2018urn,Reece:2018zvv,Lee:2018spm,Grimm:2018cpv,Buratti:2018xjt,Hebecker:2018fln,Gonzalo:2018guu,Corvilain:2018lgw,Lee:2019tst,Blumenhagen:2019qcg,Joshi:2019nzi, Marchesano:2019ifh, Font:2019cxq, Lee:2019xtm}.\footnote{A recent review on the various swampland conjectures can be found in \cite{Palti:2019pca}.}
It states that an infinite tower of modes becomes exponentially light when approaching a point that is at infinite geodesic 
distance in field space. In particular, the recent constructions \cite{Grimm:2018ohb,Grimm:2018cpv,Corvilain:2018lgw}
suggest that effective theories near such infinite distance points exhibit universal properties that can be investigated model-independently and 
quantitatively.  
As a consequence, one might thus argue that any quantum gravity conjecture should first pass its validity tests 
in the limits in field space that lie at infinite geodesic distance. We will see in this work that 
this is also a fruitful path to test the axion Weak Gravity Conjecture and uncover new mechanisms that 
can be relevant in understanding the underlying reasons for its validity. In particular, note that it was argued 
for the Swampland Distance Conjecture in \cite{Grimm:2018ohb} that the number of relevant states that have to be included 
into the effective theory has to grow with a certain rate depending on properties of the infinite distance point.
In the present context we will find the analog statement concerning the growth rate of the number of instantons 
that need to be included in the effective theory when approaching the infinite distance point. 

Our focus in this work will be on the dynamics of the 
R-R three-form axions in Type IIA string theory compactified on a Calabi-Yau threefold. These axions are 
part of the $\cN=2$ hypermultiplet field space of the effective theory. The hypermultiplet 
moduli space of Type II string theory compactification has been studied in detail, see e.g.~\cite{Alexandrov:2011va,Alexandrov:2013yva} for reviews, or \cite{Bagger:1983tt,Ferrara:1989ik,Becker:1995kb,Ooguri:1996me,deWit:1999fp,Marino:1999af,deWit:2001brd,RoblesLlana:2006ez,RoblesLlana:2006is,RoblesLlana:2007ae,Alexandrov:2008gh,Collinucci:2009nv,Alexandrov:2010ca} for some papers on this subject. 
The classical dimensional reduction shows that their kinetic terms, and hence their square axion decay constants, are proportional to the 
Hodge star metric in the Calabi-Yau manifold. 
This implies that they generally depend very non-trivially on the complex structure moduli of the Calabi-Yau threefold. 
It will be the first task of this work to descibe their  
behaviour near infinite distance points in correlation with the general classification of infinite 
distance points in the complex structure moduli space \cite{Kerr2017,Grimm:2018cpv}. We 
will find that the axion decay constants for some of the axions can, in general, 
grow to become increasingly large in the infinite distance limits. Instanton corrections  should then become 
relevant in order that the axion Weak Gravity 
Conjecture is not violated parametrically.
We suggest that these corrections stem from D-brane states and identify candidate
D2-brane instanton states wrapped on three-cycles of the Calabi-Yau manifold that 
non-trivially modify the effective theory at such points. Our construction follows the 
ideas of \cite{Grimm:2018ohb, Grimm:2018cpv} to characterize a tower of 
D3-brane states wrapped on three-cycles that gives the relevant particles for the Swampland 
Distance Conjecture in Type IIB compactifications. 

The axion Weak Gravity Conjecture has already been investigated in various ways, see for instance \cite{Cheung:2014vva,delaFuente:2014aca,Rudelius:2015xta,Montero:2015ofa,Brown:2015iha, Bachlechner:2015qja, Brown:2015lia,Junghans:2015hba,Heidenreich:2015wga,Palti:2015xra,Ibanez:2015fcv, Hebecker:2015zss,Heidenreich:2015nta, Long:2016jvd,  Heidenreich:2016aqi, Hebecker:2018fln,Andriolo:2018lvp,Marchesano:2019ifh}. One of the main challenges in addressing the axion Weak Gravity Conjecture arises when  
dealing with higher-dimensional field spaces. In this case one does not expect a simple 
direct link between the instanton actions and the axion decay constants. In fact, the kinetic 
term of the axions will generally be given in terms of a non-diagonal field-dependent metric 
and instanton actions will generally not align with any diagonalization attempt. A related issue 
in each higher-dimensional setting is the fact that we have to deal with path dependence when 
approaching an infinite distance point. In particular, the kinetic term for a certain axion might grow 
along one specific path, but might stay finite along another. It is one of our main tasks 
to address these general issues for the Type IIA setting under consideration. We will argue 
that there is a natural basis for the axions that is adapted to the infinite distance locus under 
consideration. More precisely, this special basis will arise from the fact that we can non-trivially  
associate an $\mathfrak{sl}(2,\bbC)^n$-algebra to each infinite 
distance locus reached by sending $n$
coordinates into a limit \cite{Schmid,CKS}. This algebra acts on the three-forms defining the 
axion basis and thus splits the axion space into subspaces. 
We can then generally determine the growth of these subspaces when reaching the infinite 
distance point. This allows us to focus on the set of axions that have decay constants 
that grow parametrically in a certain path-independent way. 
 
This paper is structured as follows. In section \ref{sec:infinite_distance_points} we 
first review the classification of infinite distance points in complex structure moduli space.  
We then introduce a special real three-form basis adapted to the $\mathfrak{sl}(2,\bbC)^n$-algebra associated to 
the infinite distance limit and discuss the growth of the associated Hodge 
metric. In section \ref{sec:SDC} we then recall some of the recent insights about the Swampland 
Distance conjecture for Calabi-Yau moduli spaces \cite{Grimm:2018ohb, Grimm:2018cpv} and adapt the 
presentation to the special three-form basis of section \ref{sec:infinite_distance_points}, whereafter we will generalize the stability argument presented in \cite{Grimm:2018ohb} to multi-variable settings. The axion Weak Gravity Conjecture will then be addressed in section \ref{WGCaxions}. We identify 
a candidate tower of D2-brane instantons which prevents the parametric violation of this 
conjecture.

\noindent  
\textit{Note added:} While we were in the process of writing up this paper the reference \cite{Marchesano:2019ifh} appeared, which 
has some overlap with our work, but is in many respects complementary. On the one hand, our Type IIA 
treatment is general and does not focus on particular infinite distance limits, while \cite{Marchesano:2019ifh} focuses on specific 
limits in the large volume regime. On the other hand \cite{Marchesano:2019ifh} gives an interesting discussion on the modifications 
of the effective theories, while we will not address the modifications of the effective theory in the present work.

\section{Infinite distances in Calabi-Yau moduli spaces} \label{sec:infinite_distance_points}

The aim of this work is to study the couplings of R-R axions in the four-dimensional effective 
theory. To motivate the use of the mathematical tools introduced in this section, let us first briefly 
recall the relevant structures in Type IIA 
Calabi-Yau compactifications. In this case we consider the axions $\xi^\cI$ arising 
from expanding the R-R field $C_3$ into a 
basis $\gamma_\cI$ of the third cohomology group $H^3(Y_3,\bbR)$ via
\beq \label{C3-expand}
    C_3 =   \xi^\cI \gamma_\cI\ .
\eeq
The four-dimensional kinetic terms are readily derived to be \footnote{Let us stress that in a Calabi-Yau threefold reduction of Type IIA the axions 
reside in a quaternionic field space and have additional kinetic terms. These terms play no important role for the considered limits discussed in section \ref{WGCaxions}, and are absent in Type IIA orientifold truncations \cite{Grimm:2004ua,Grimm:2007hs}.}
\beq \label{axion-metric}
  \cL_{\rm kin} = G_{\cI \cJ} \ \partial_\mu \xi^\cI \partial^\mu \xi^\cJ\ , \qquad G_{\cI \cJ} = \frac{1}{2} e^{2D} \int_{Y_3} \gamma_\cI \wedge * \gamma_\cJ  \ ,
\eeq
where $e^D$ is the four-dimensional dilaton and $*$ is the Hodge star of the Calabi-Yau threefold. Note that 
the metric $G_{\cI \cJ}$, crucial in defining the axion decay constants as we discuss below, 
non-trivially depends on the complex structure moduli through $*$.

In this section we will discuss the techniques needed to analyze limits in the moduli space of Calabi-Yau manifolds
in which the axion metric $G_{\cI \cJ}$ grows. We will not be able to introduce the complete mathematical theory relevant to answer these questions, 
but rather constrain ourselves to stating some of its main results from \cite{Schmid,CKS,Kashiwara,wang1,Kerr2017}. For a proper mathematical review on the foundations of this subject see e.g.~\cite{AST_1989__179-180__67_0}. More details are also provided in \cite{Grimm:2018cpv}.

\subsection{On the geometry of the complex structure moduli space}

In order to set the stage for the later discussions, let us first recall some basic properties of the 
complex structure moduli space $\mathcal{M}^{\rm cs}$. This moduli space is spanned by the complex structure deformations 
that preserve the Calabi-Yau property and has complex dimension $h^{2,1}$, where 
$h^{p,q}=\text{dim}(H^{p,q}(Y_3,\mathbb{C}))$ are the Hodge numbers of $Y_3$. 
Since $\mathcal{M}^{\rm cs}(Y_3)$ admits a special K\"ahler structure its 
metric can be derived from a K\"ahler potential $K$.  Let us introduce 
local coordinates $z^I$ with $I=1, \ldots, h^{2,1}$. Furthermore, recall that the Calabi-Yau threefold 
$Y_3$ admits a unique  $(3,0)$-form $\Omega(z)$ that varies holomorphically in these coordinates $z^I$. 
 $\Omega$ can be used to define the metric on $\mathcal{M}^{\rm cs}$ via the K\"ahler potential
\begin{equation} 
    K(z,\bar z) = -\log \Big[  i \int_{Y_3} \Omega \wedge \bar{\Omega} \Big] \ .
\end{equation}
Next we introduce a real, integral basis $\gamma_{\mathcal{I}}$ for $H^3(Y_3,\mathbb{Z})$, with $\mathcal{I}=1,\ldots, 2h^{2,1}+2$. This allows us to decompose $\Omega$ in its periods $\Pi^{\mathcal{I}}$ as 
\begin{equation}
        \Omega(z) = \Pi^{\mathcal{I}}(z)\gamma_{\mathcal{I}}\ .
\end{equation}
Furthermore, we can construct a skew-symmetric product $\langle \cdot, \cdot \rangle$ on $H^3(Y_3,\mathbb{C})$, or component-wise an anti-symmetric pairing matrix $\eta$, given by \footnote{Note that the skew-symmetric product in \cite{CKS}, and also \cite{Grimm:2018ohb, Grimm:2018cpv}, was denoted by 
$S(v,u)$ and differs by a minus sign from the definition used here.}
\begin{equation} \label{def-<>}
\langle v,w \rangle =  \int_{Y_3} v \wedge w, \qquad \eta_{\mathcal{I}\mathcal{J}} = - \langle \gamma_{\mathcal{I}}, \gamma_{\mathcal{J}} \rangle ,
\end{equation}
with $v,w \in H^3(Y_3,\mathbb{C})$. Then we can express the K\"ahler potential as
\begin{equation} \label{K-Pi}
    K = -\log\big[ i \langle  \mathbf{\Pi}  , \bar{\mathbf{\Pi}}\rangle \big]=-\log \Big[ i  \bar{\Pi}^{\mathcal{I}} \eta_{\mathcal{I}\mathcal{J}} \Pi^{\mathcal{J}} \Big] .
\end{equation}
Note that due to the skew-symmetry of $\langle \cdot, \cdot \rangle$ one can choose a symplectic basis 
$\gamma_{\mathcal{I}}=(\alpha_{\hat{K}},\beta^{\hat{L}})$ with $\hat{K},\hat{L}=0,\ldots , h^{2,1}$. This basis satisfies the following properties
\begin{equation} \label{sympl_basis}
    \langle \alpha_{\hat{K}} ,  \beta^{\hat{L}} \rangle = \delta_{\hat{K}}^{\hat{L}}, \qquad \langle \alpha_{\hat{K}}, \alpha_{\hat{L}} \rangle = \langle \beta^{\hat{K}}, \beta^{\hat{L}} \rangle = 0.
\end{equation}
As we will discuss in detail in section \ref{special_basis} there is a very special choice of such a symplectic basis associated to the considered point in 
moduli space when analyzing asymptotic limits. 


\subsection{Limits in the complex structure moduli space}

We next discuss the relevant limits in this complex structure moduli space $\mathcal{M}^{\rm cs}$. 
It will turn out that the metric $G_{\cI \cJ}$ of the axions \eqref{axion-metric} can only grow unboundedly if we 
approach a point on in $\cM^{\rm cs}$ at which the Calabi-Yau manifold $Y_3$ degenerates. Well-known 
examples of such degeneration points are the conifold point or the large complex structure point, but our analysis will be completely 
general and include also higher-dimensional degeneration loci. 
It can be shown that one can blow-up $\cM^{\rm cs}$ in such a way that the subspaces at 
which $Y_3$ degenerates can locally be described as the vanishing locus of $n$ coordinates $z^1=...=z^n=0$.\footnote{This 
equation describes the intersection of $n$ divisors in the blown-up $\cM^{\rm cs}$.} 
Instead of working with the $z^I$
we will introduce new coordinates $t^i = \frac{1}{2 \pi i} \log z^i $ such that the  limits of interest are given by
\beq \label{t-limit}
   t^1 \,, \ldots\, , t^n \ \rightarrow \ i\infty\ , \qquad \zeta^\kappa \ \text{fixed}\ ,
\eeq
where $ \zeta^\kappa$ are the coordinates that are not taken to a limit. 

\begin{figure}[h!]
\vspace*{.5cm}
\begin{center} 
\includegraphics[width=8cm]{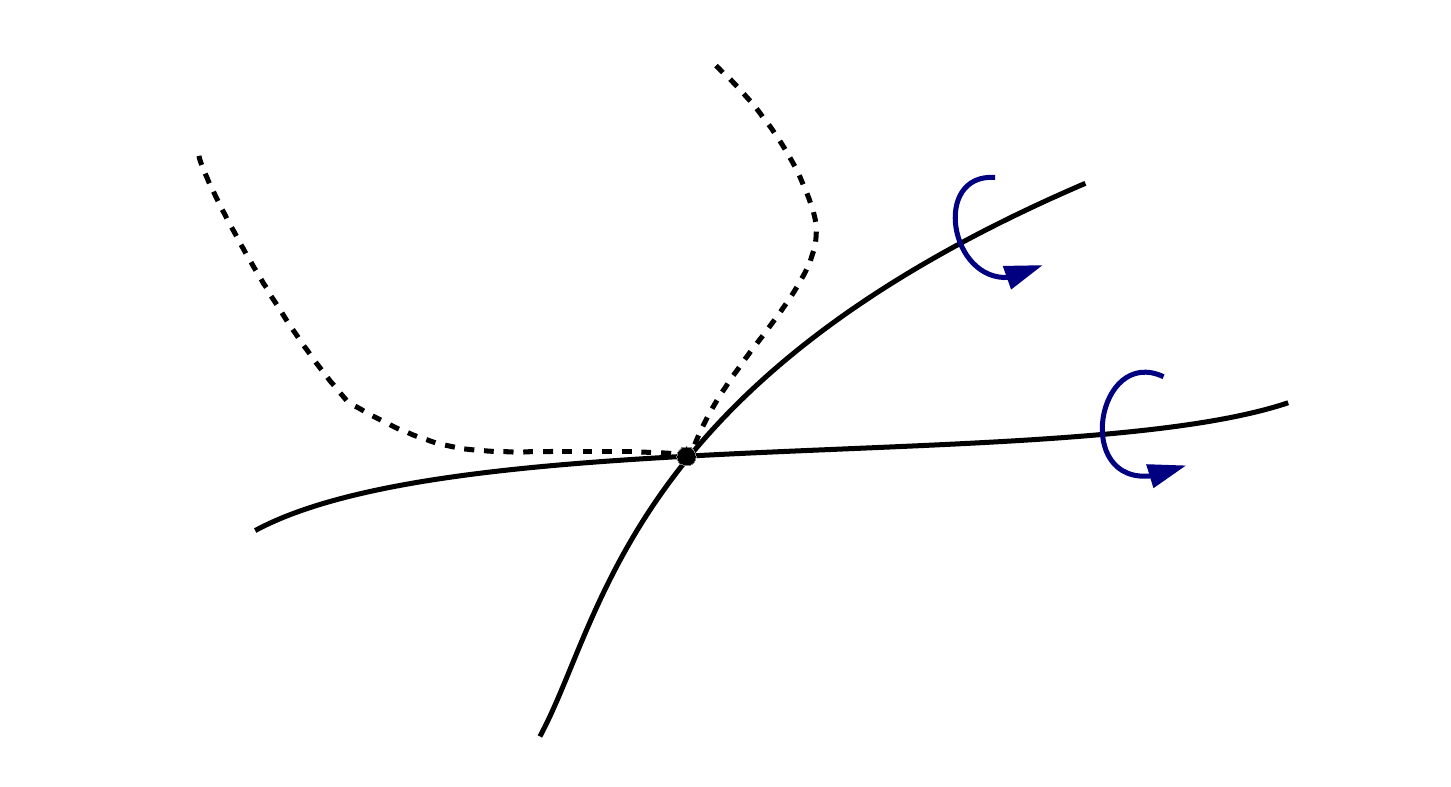} 
\vspace*{-1cm}
\end{center}
\begin{picture}(0,0)
\put(285,90){$z_1 = 0$}
\put(317,55){$z_2=0$}
\put(258,97){$N_1$}
\put(268,58){$N_2$}
\put(222,85){$ \gamma_{2}$}
\put(172,60){$ \gamma_{1}$}
\end{picture}
\caption{Two paths $\gamma_1,\gamma_2$ towards the limiting point $t^1\rightarrow i \infty\ (z_1 = 0)$ and 
$t^2 \rightarrow i \infty\ (z_2 = 0)$, respectively. 
The paths can lie in different growth sectors, since either the growth of $y^1$ or $y^2$ dominate in these cases. 
We also indicated the associate log-monodromy matrices $N_1,N_2$.} \label{intersecting_div}
\end{figure}
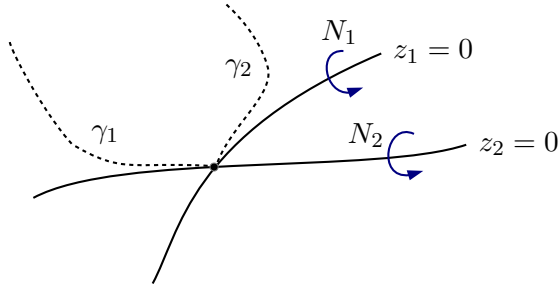

The growth of the axion metric $G_{\cI \cJ}$ in the limits \eqref{t-limit} will in general depend on the precise path we are taking 
to these limiting points. The mathematical machinery we intend to use does provide us with general growth estimates 
in case we first divide the space into sectors, so-called growth sectors, and then demand that the considered path lies within 
one of such growth sectors at least 
for sufficiently large $y^i = \Im t^i$. One of such growth sectors is given by 
\begin{equation}\label{eq:growthsector}
    \mathcal{R}_{12\cdots n} = \Big\{t^i=x^i+iy^i \Big|\, \frac{y^1}{y^2}>\lambda \, ,\ \ldots, \frac{y^{n-1}}{y^n}>\lambda , y^n>\lambda, \ x^i < \delta \Big\}\ ,
\end{equation}
where we can chose arbitrary positive $\lambda,\delta$. Other growth sectors can be obtained by the same expression but with permuted 
$y^i$. Clearly, if a path lies within one of these sectors we can relabel the $y^i$ such that the respective sector is given by \eqref{eq:growthsector}.
In figure \ref{intersecting_div} we 
illustrate two paths for the case that two $t^i$ are send to the limit. Let us stress that the requirement that the 
considered path resides in one growth sector introduces a mild path dependence into our analysis, 
since we exclude paths which are so complicated that they always pass through multiple sectors.

It is a famous mathematical result \cite{Schmid} that the limiting behaviour of the periods $\mathbf{\Pi}$, and also the metric $G_{\cI\cJ}$, 
crucially depends on the monodromy matrix $T_i$ associated to the $t^i = i \infty$ point. This monodromy matrix appears 
if one asks how $\mathbf{\Pi}$ transforms under $t^i \to t^i - 1$ for some index $i$, 
i.e.~one has 
\beq \label{monodromy}
   \mathbf{\Pi}(\ldots, t^i - 1,\ldots)=T_i \mathbf{\Pi}(\ldots, t^i,\ldots)\ .
\eeq 
The monodromy matrices $T_i$ turn out to possess several very useful properties. Firstly, all  $T_i$ 
associated to the limit \eqref{t-limit} commute with each other. Secondly, if a $T_i$ possesses a non-trivial unipotent part, it 
defines a nilpotent matrix \footnote{In the following we will assume that we have transformed the variables $z^i$ and $t^i$, such that 
only the uni-potent part of $T_i$ is relevant in the transformation \eqref{monodromy}. This procedure causes us to lose some of the information about the monodromies of orbifold singularities, but the aspects crucial to the infinite distances are retained.}
\beq
    N_i = \log T_i \ . 
\eeq
The $N_i$ also form a commuting set of matrices and one has $\eta N_i = - N_i^T \eta$. 
The nilpotent orbit theorem of \cite{Schmid} allows us to express $\mathbf{\Pi}$ in terms of the nilpotent matrices \footnote{Note that this 
statement is true up to an overall holomorphic rescaling of $\mathbf{\Pi}$. Such rescalings yield to a K\"ahler transformation of 
$K$ given in \eqref{K-Pi}. Unless otherwise indicated the following discussion is invariant under such rescalings.}
\begin{equation} \label{nilp-orbit_approx}
    \mathbf{\Pi} = e^{- t^i N_i}\mathbf{A}(e^{2\pi i t^i},\zeta) =e^{-t^i N_i} \big(\mathbf{a}_0(\zeta)+\mathcal{O}(e^{2\pi i t^i},\zeta) \big) ,
\end{equation}
where we sum in the exponential over $i=1,...,n$. Here 
$\mathbf{a}_0$ is a holomorphic function in the coordinates that are not send to a limit \eqref{t-limit}. 
Note here that the exponential yields a polynomial in $t^i$, since the $N_i$ are nilpotent matrices. The 
important statement is that the vector $\mathbf{A}$ is holomorphic in $z^A = e^{2\pi i t^A}$ and $\zeta$, 
which allows for the above 
expansion with leading term $\mathbf{a}_0$. This vector $\mathbf{a}_0$ determines the asymptotic behavior of $\mathbf{\Pi}$ in the limit 
\eqref{t-limit}, since the other terms will be suppressed if we take $\text{Im}\, t^i$ to be large. 
$\mathbf{a}_0$ naturally defines a so-called nilpotent orbit given by 
\beq \label{Pinil}
    \mathbf{\Pi}_{\rm nil} ( t,\zeta)= e^{-t^i N_i} \mathbf{a}_0(\zeta)
\eeq
This nilpotent orbit is the starting point for our analysis of the asymptotic regions in $\cM^{\rm cs}$. 

\subsection{Classifying infinite distance limits} \label{classify_limits}

The information captures by $\mathbf{\Pi}_{\rm nil}$ or $(N_i,\mathbf{a}_0)$ can be used to classify infinite distance limits. 
Recall that the distance between two points $P,Q$ along a path $\gamma$ is determined by the integral 
\beq
    d_{\gamma}(P,Q) = \int_\gamma \sqrt{g_{k \bar l} \dot t^k \dot {\bar t}^l} ds
\eeq
where in the complex structure moduli space one has to use the metric $g_{k \bar l} = \partial_{t^k} \partial_{\bar t^l} K$ determined 
from \eqref{K-Pi}. In order that the geodesic distance between points $P,Q$ is infinite, every path $\gamma$ between $P,Q$ has to be 
of infinite lengths. This can only potentially happen if one of the points, say $P$, is located at one of the loci $t^1=...=t^n = i \infty$.  
However, not every such locus is at infinite distance. In fact, using the nilpotent orbit \eqref{Pinil} and the properties of 
$\mathbf{a}_0$ one shows that \cite{wang1}
\beq \label{Pinfinite}
   P\ \text{at infinite distance}: \quad N_{(n)} \mathbf{a}_0 \neq 0 \ , 
\eeq
as discussed in detail in \cite{Grimm:2018cpv}. Here we have defined $ N_{(n)} = N_1 + ... + N_n$, but stress that 
every linear combination of the $N_i$ with positive coefficients could equally be used. 
%
%

It was shown in \cite{Grimm:2018ohb,Grimm:2018cpv} that it is crucial to actually distinguish several cases of 
infinite distance limits. In order to do that one needs to analyze the the properties 
of $N_{(n)}$ and $\eta$. In fact, one can analyze the occurring singularity for 
any step sending $t^1,...,t^i \rightarrow i \infty$, for $i=1,...,n$, associating an $N_{(i)}=N_1+...+N_i$. 
 For each such pair $(N_{(i)},\eta)$ one finds one of $4 h^{2,1} $ types of limiting behaviours denoted by  \cite{Kerr2017}
\beq \label{all_limits}
\text{I}_a \ , \quad  \text{II}_b \ , \quad  \text{III}_c\ , \quad  \text{IV}_d\ ,
\eeq
where $a,b,c,d$ are indices with index ranges listed in table \ref{sing_type}. The table 
also gives the rules that allow to associate the types to $(N_{(i)},\eta)$. Let us stress that each of this limits 
corresponds to making $Y_3$ singular and \eqref{all_limits} provides a classification of all allowed singularities.  
\begin{table}[!ht]
	\centering
	\begin{tabular}{| c |c| l | lcc| r|} \hline
		\multirow{2}{*}{Type} & \multirow{2}{*}{$d_i$} &  \multirow{2}{*}{Index range}  &\multicolumn{3}{c|}{  \rule[-0.2cm]{0cm}{0.7cm} \hspace*{-1.5em} rank of} & \multirow{2}{*}{eigenvalues of $\eta N_{(i)}$}  \\[1pt] 
		\rule[-0.3cm]{0cm}{0.6cm} && &$N_{(i)}$ &$N_{(i)}^2$&$N_{(i)}^3$ & \\ \hline \hline
		\rule[-0.2cm]{0cm}{0.7cm} I$_a$ & 0 &  $a=0,\ldots,h^{2,1}$ &$a$ & 0& 0 &$a$ negative\\
		\rule[-0.2cm]{0cm}{0.7cm} II$_b$ &1 & $b=0,\ldots,h^{2,1}-1$ &$2+b$ & 0& 0 & 2 positive, $b$ negative\\
		\rule[-0.2cm]{0cm}{0.7cm} III$_c$ &2 & $c= 0,\ldots,h^{2,1}-2$  &$4+c$ & 2& 0 & not needed\\
		\rule[-0.2cm]{0cm}{0.7cm} IV$_d$ &3 &  $d = 1,\ldots,h^{2,1}$ &$2+d$ & 2& 1 & not needed\\		\hline 
	\end{tabular}
	\caption{Classification of limits in the complex 
		moduli space.}
	\label{sing_type}
\end{table}

One can show that the type at each step $i$ also determines the highest integer $d_i$ such that
\begin{equation} \label{def-di}
   N_{(i)}^{d_i} \mathbf{a}_0 \neq 0 \ , 
\end{equation}
with $d_{i}=0,1,2,3$ for the four types I, II, III, and IV, respectively. We have included these labels 
in table \ref{sing_type}. 
Since we are interested in limits that lie at infinite distance, we can use \eqref{Pinfinite}
to infer
\beq
  \text{infinite distance limits:} \quad (N_{(n)},\eta)\ \ \text{is of type}\ \ \text{II}_b \, , \  \text{III}_c  \ \text{or}\  \text{IV}_d\ . 
\eeq
It will be these limits in which we will study the behaviour of the axion metric $G_{\cI \cJ}$ 
given in \eqref{axion-metric}. 

\subsection{A special three-form basis and the $\mathfrak{sl}(2)$-splitting} \label{special_basis}

Having classified the infinite distance limits in $\cM^{\rm cs}$ we 
next want to connect this information with the axion metric \eqref{axion-metric} for the axions $\xi^\cI$ 
arising in the expansion \eqref{C3-expand}. In order to do that it turns out to be very useful to 
introduce a special basis $\gamma_\cI$ for $H^3(Y_3,\mathbb{R})$, which is adapted to the 
limiting locus that we approach. More precisely, the basis will depend on the following set of 
data: 
\begin{itemize} 
\item[(1)] the monodromy matrices $N_i$ relevant for the considered limit; 
\item[(2)] the limiting vector $\mathbf{a}_0$ relevant for the considered limit;
\item[(3)] the growth sector \eqref{eq:growthsector} in which the considered path resides. 
\end{itemize}
The rough idea is to split up $H^{3}(Y_3,\bbR)$ into a direct sum of smaller subspaces whose 
elements have a particular growth in the fields $y^i=\Im t^i$ approaching infinity. Furthermore, one 
finds a `limiting Hodge metric', in which these spaces are orthogonal. At first, this seems like an 
impossible task, since even the approximate periods \eqref{Pinil} contain contain numerous 
mixed terms due to the general form of the $N_i$ and $\mathbf{a}_0$. However, there is the 
famous formalism of \cite{Schmid,CKS} that allows to systematically approach this problem. 

The in-genius idea of \cite{CKS} is to reformulate this structure such that it non-trivially `decomposes'  
into $\mathfrak{sl}(2,\mathbb{C})$-blocks that decouple in a well-defined sense. 
More precisely, \cite{CKS} constructs a set of $n$ mutually commuting $\mathfrak{sl}(2,\mathbb{C})$-triples 
acting on $H^3(Y_3,\bbR)$ form the above local data (1)-(3).  We will not describe the steps to 
actually perform this construction, but refer the reader to \cite{Grimm:2018cpv} for a detailed review and the study of an explicit example. 
Let us simply assert that we went through the relevant steps and introduce the 
\beq 
    \text{commuting}\ \mathfrak{sl}(2,\mathbb{C})\text{-triples}: \qquad (N_i^-,N_i^+,Y_i)\ , \quad i=1,...,n\ .
 \eeq 
These triples satisfy the standard commutation relations $[Y_i,N_i^{\pm} ]= \pm 2N_i^{\pm}$ and $[N_i^+,N_i^-]=Y_i$. 
We can now use these triples to split $H^3(Y_3,\mathbb{R})$ into eigenspaces. Let us introduce
\beq \label{split-Vell}
   \boxed{\quad \rule[-.4cm]{0cm}{1.0cm} H^3(Y_3,\mathbb{R}) = \bigoplus_{\underline{\ell}\in \cE} V_{\underline{\ell}}\ ,\qquad  \underline{\ell}=(\ell_1, \ldots, \ell_n)\ ,\quad }
\eeq
where $\ell_i \in \{0,...,6 \}$ are integers representing the eigenvalues of $Y_{(i)}=Y_1 + \ldots + Y_i$, i.e.
\beq\label{eigenvalue}
  v_{\underline{\ell}} \in V_{\underline{\ell}}: \qquad Y_{(i)} v_{\underline{\ell}} = (\ell_i - 3) v_{\underline{\ell}}\ .
\eeq
We have denoted by $\cE$ the set of all possible vectors $\ell$ labelling non-trivial $V_{\underline{\ell}}$ and 
collecting all eigenvalue combinations of $(Y_{(1)},...,Y_{(n)})$. Let us stress 
that the range of $\ell_i$ labelling non-empty $V_{\underline{\ell}}$ is correlated with the type of limit associated to $(N_{(i)},\eta)$ when sending $t^1,...,t^i \rightarrow i \infty$ as listed in table \ref{sing_type}.
In fact, one finds 
\begin{align}
    \text{I,\ II} \, :\    \ell_i =2,...,4 \ ,  \quad
     \text{ III} \, :\    \ell_i   = 1,...,5\ ,  \quad
   \text{ IV} \, :\     \ell_i   = 0,...,6\ .  
\end{align}
Note that using the 
fact that the singularity can only increase sending more $t^i$ to the limit we find that the 
range of $\ell_i$ successively increases with $i$.

One can derive several interesting properties of the vector spaces $V_{\underline{\ell}}$. 
Most important for us is the fact that 
\beq \label{dim-eq}
   \text{dim} V_{\underline \ell} =   \text{dim} V_{\underline{6} - \underline \ell} \ ,
\eeq
where we abbreviated $\underline{6} = (6,...,6)$. This implies that we can one-to-one identify 
a basis vector of $V_{\underline \ell}$ and $V_{\underline{6} - \underline \ell} $.
Furthermore, these spaces $V_{\underline{\ell}}$ satisfy certain orthogonality relations, as follows from
\begin{equation}
(r_i-3) \langle V_{\underline{\ell}}, V_{\underline{r}} \rangle =  \langle V_{\underline{\ell}}, Y_{(i)} V_{\underline{r}} \rangle 
=- \langle Y_{(i)} V_{\underline{\ell}},  V_{\underline{r}} \rangle
= (3-\ell_i) \langle V_{\underline{\ell}}, V_{\underline{r}} \rangle\, ,
\end{equation}
by using \eqref{eigenvalue} and that  $\langle \cdot,Y_{(i)} \cdot \rangle  = - \langle Y_{(i)} \cdot, \cdot \rangle  $. Namely, it implies that the product between these spaces can only be non-zero if $\ell_i+r_i = 6$. And since this should hold for all $i$, the vector spaces $V_{\underline{\ell}}$ satisfy the 
orthogonality property 
\beq \label{V-orth}
    \langle V_{\underline{\ell}}, V_{\underline{\ell}'} \rangle =  0 \quad \text{unless} \quad \underline{ \ell}+\underline{\ell}' = \underline{6}\ . 
\eeq
It should now be clear from \eqref{dim-eq}, \eqref{V-orth}, and the skew-symmetry of the inner product $\langle \cdot,\cdot\rangle$
that the $V_{\underline{\ell}}$ define naturally a special symplectic basis with the properties \eqref{sympl_basis}. 

Importantly the limiting vector $\mathbf{a}_0$ turns out to not generally fall into one of the spaces $V_{\underline{\ell}} \otimes \bbC $ of the 
splitting \eqref{split-Vell}. However, as shown in \cite{CKS} and discussed in detail in \cite{Grimm:2018cpv}, there are always two 
real matrices $\zeta',\delta$ which rotate $\mathbf{\tilde a_0} = e^{\zeta'} e^{- i \delta} \mathbf{a}_0$, such that 
\beq \label{location_a0}
     \mathbf{\tilde a_0}\  \in\ V_{\underline{3}+\underline{d}} \otimes \bbC  \ , \quad \underline{d} = (d_1,...,d_n)\ , 
\eeq
with $d_i$ defined in \eqref{def-di}. Crucially, this construction is such that one also finds that the complex conjugate of 
$  \mathbf{\tilde a_0}$ lies in $V_{\underline{d}}$, such that 
\beq  \label{location_RIa0}
    \R\, \mathbf{\tilde a_0}\, , \ \I\, \mathbf{\tilde a_0}\  \in\ V_{\underline{3}+\underline{d}} \ , \quad \underline{d} = (d_1,...,d_n)\ .
\eeq
 The vector $ \mathbf{\tilde a_0}$ generally depends on the remaining coordinates $\zeta^\kappa$
not taken to a limit in \eqref{t-limit}. It can be used to define the so-called Sl(2)-orbit 
\beq \label{Sl2-orbit}
  \mathbf{\Pi}_{\rm Sl(2)} (y,\zeta)= e^{- i y^i N^-_i} \mathbf{\tilde a}_0(\zeta)\ .
\eeq
This orbit asymptotically approximates the nilpotent orbit \eqref{Pinil} in the limit 
$\frac{y^1}{y^2},...,\frac{y^{n-1}}{y^n}, y^n \rightarrow \infty$ and $x^i= 0$.

It should be stressed that many key properties of our later constructions are contained in this very non-trivial 
$\mathfrak{sl}(2)$-split \eqref{split-Vell} of $H^{3}(Y_3,\bbR)$. 
One of these properties, namely the growth of the Hodge metric of this basis, we will discuss next.

\subsection{Asymptomatic behavior of the Hodge norm}\label{asymp_Hodge_norm}

One of the remarkable applications of the $\mathfrak{sl}(2)$-splitting, which we introduced in section~\ref{special_basis}, is 
to obtain an asymptotic expression for the Hodge metric that appears, for example, in the definition \eqref{axion-metric} of the axion metric $G_{\cI \cJ}$. 

Let us first introduce some notation and define the Hodge norm of a three-forms $\mathbf{v} \in H^3(Y_3,\mathbb{C})$ by
\begin{equation} \label{def-Hodgenorm}
\| \mathbf{v} \|^2 = \langle  \mathbf{\bar v}, \ast \mathbf{v} \rangle = \int_{Y_3}  \mathbf{\bar v} \wedge \ast \mathbf{v}\ .
\end{equation}
As stressed above, the Hodge star $*$ in general depends very non-trivially on the complex structure moduli. In order to make this 
dependence explicit, one can decompose $\mathbf{v}$ into its components in $H^{p,q}(Y_3,\bbC)$, with $p+q=3$. The individual
components can then be expressed in terms of the period vector $\mathbf{\Pi}$ and its K\"ahler-covariant derivatives. 
It turns out that one can control the asymptotic of the periods $\mathbf{\Pi}$ in the limits \eqref{t-limit} which then leads to 
an asymptotic expression for the Hodge metric.\footnote{This again non-trivially applies the $\mathfrak{sl}(2)$-splitting introduced 
in the previous subsection.} 

To make the asymptotic form of the metric explicit, we first introduce a limit Hodge norm \cite{CKS,AST_1989__179-180__67_0}
\beq \label{def-norminf}
     \| \mathbf{v} \|_\infty^2 = \langle  \mathbf{\bar v}, \ast_\infty \mathbf{v} \rangle \ .
\eeq
While we will not define $\ast_\infty$ in any detail, let us record some of its properties. Firstly, $\|\cdot \|_\infty$ is finite in the limit \eqref{t-limit}, since $\ast_\infty$ does no longer depend on the fields $t^1,...,t^n$.\footnote{Note that $*_\infty$ can still depend on the coordinates $\zeta$
and hence become singular if these are sent into special limits.} It is adapted to the splitting of 
the vector space $H^{3}(Y_3,\bbR)$ introduced in \eqref{split-Vell} via the orthogonality relations 
\beq \label{V*V}
   \langle  V_{\underline{\ell}}, \ast_\infty V_{\underline{\ell}'} \rangle  = 0 \quad \text{unless}\quad \underline{\ell} = \underline{\ell}'\ .
\eeq
The norm \eqref{def-norminf} can be used to give an asymptotic expression for the 
original Hodge norm \eqref{def-Hodgenorm} in the limit \eqref{t-limit}. Decomposing 
a general real three-form $\mathbf{u}$ into its components $\mathbf{u}_{\underline{\ell}} \in V_{\underline{\ell}} $ one 
has \cite{Kashiwara,CKS}
\begin{equation} \label{general-norm-growth}
 \boxed{\quad \rule[-.4cm]{0cm}{1.1cm}  \| \mathbf{u} \|^2\ \sim\ \sum_{\underline{\ell} \in \cE} \Big( \frac{y^1}{y^2}\Big)^{\ell_{1}-3} ... \Big(\frac{y^{n-1}}{y^n} \Big)^{\ell_{n-1}-3}
 (y^n)^{\ell_{n}-3}\, \|\mathbf{u}_{\underline{\ell}}\|^2_\infty \ , \qquad \mathbf{u} = \sum_{\underline{\ell} \in \cE} \mathbf{u}_{\underline{\ell}}\ . \quad}
\end{equation}
This expression will be our main tool in the rest of this paper to evaluate the 
growth of the axion metric $G_{\cI \cJ}$ and the associated D2-brane instanton actions.


\section{The Swampland Distance Conjecture for Calabi-Yau moduli}\label{sec:SDC}
In this section we revisit the recent constructions of \cite{Grimm:2018cpv, Grimm:2018ohb} that
provided strong evidence for the validity of the Swampland Distance Conjecture (SDC) for all infinite distance limits in 
$\cM^{\rm cs}(Y_3)$. The crucial observation of these works is, that one can 
relate the classification of infinite distance points, recalled in section \ref{classify_limits}, with the existence of a tower of D3-branes wrapped on three-cycles of $Y_3$ with masses becoming 
exponentially light when approaching the infinite distance points. One of the main tasks in establishing such a picture is
the search for suitable three-cycles that can host such states. While it will not add much to the strategy presented 
in \cite{Grimm:2018cpv, Grimm:2018ohb}, 
we will reformulate and generalize the statements using the special basis introduced in section \ref{special_basis}. 
This reformulation turns out to be an elegant 
way of stating the findings and will serve as a prelude to section \ref{WGCaxions}, where we will consider axion decay constants and
Euclidean D2-branes wrapping three-cycles of $Y_3$. Furthermore, we will generalize the stability properties of the D3-brane states of \cite{Grimm:2018ohb}, where they studied the one-parameter setting, to the multi-parameter infinite distance limits considered in \cite{Grimm:2018cpv}, which will play an important role in the test of the axion Weak Gravity Conjecture in section \ref{WGCaxions}.

\subsection{Construction of the D3-brane states}\label{sec:constructionofstates}

Let us begin by introducing the necessary basic properties of the three-cycles that can host the 
D3-brane states required to satisfy the SDC. 
In order to specify the state that we obtain from wrapping a D3-brane on a three-cycle 
of $Y_3$, we will give its Poincar\'e dual three-form $\mathbf{Q} \in H^3(Y_3,\mathbb{R})$. Since we are mainly interested 
in the mass of this state we will not discuss the quantization of $\mathbf{Q}$ in the following.\footnote{We note that many 
aspects of the structure introduced here can be generalized over $\mathbb{Q}$.} Since we are performing 
a Calabi-Yau compactification the four-dimensional theory is an $\cN = 2$ supergravity theory.
Our aim is to considere candidate charges $\mathbf{Q}$ that correspond to BPS states. This non-trivial assertion will 
have severe consequences, as we discuss below. In the following 
we will first establish the conditions on $\mathbf{Q}$ in order that the 
corresponding D3-branes state becomes light in an infinite distance limit $t^1,...t^n \rightarrow i \infty$
introduced in \eqref{t-limit}. 

Given a BPS D3-brane state with charge $\mathbf{Q}$ we can compute its mass by evaluating its central charge 
using $M(\mathbf{Q}) = |Z(\mathbf{Q})|$, with the central charge given by
\begin{equation}\label{central_charge}
Z(\mathbf{Q}) = \frac{\langle  \mathbf{Q} , \mathbf{\Pi}\rangle }{\| \mathbf{\Pi} \|}\, ,
\end{equation}
where $\mathbf{\Pi}$ are the periods appearing in \eqref{K-Pi} and $\langle \cdot , \cdot \rangle$, $\|\cdot \|$ were defined in \eqref{def-<>}, \eqref{def-Hodgenorm}.
We are interested in finding the charge vectors $\mathbf{Q}$ such that the states become massless in the infinite distance limit. Since $\ast \mathbf{\Pi} = -i \mathbf{\Pi}$ 
we can apply Cauchy-Schwarz inequality to find the following upper bound on the mass of the state
\begin{equation}
M(\mathbf{Q}) =  \frac{|\langle  \mathbf{Q} ,\ast \mathbf{\Pi}\rangle | }{\| \mathbf{\Pi} \|} \leq \| \mathbf{Q} \| \, .
\end{equation}
This implies that a sufficient condition for becoming light in the limit \eqref{t-limit} is that $\| \mathbf{Q} \| \rightarrow 0$. 
Classifying in a higher-dimensional moduli space the states that admit such a behaviour for a general path approaching 
the infinite distance point is clearly challenging. However, as 
we will see in the following the machinery introduced in section \ref{sec:infinite_distance_points} allows us to do this for all paths that reside in a single 
growth sector.  

Let us now consider a path with \eqref{t-limit} that approaches an infinite distance point and eventually resides in the growth sector \eqref{eq:growthsector}. 
In this case we can apply the growth result \eqref{general-norm-growth}. The requirement that $\mathbf{Q}$ behaves as $\| \mathbf{Q} \| \rightarrow 0$ 
along each such path, leads us to define the vector space 
\beq \label{def-Vlight}
  V_{\rm light} =  \bigoplus_{\underline{\ell} \in \cE_{\rm light}} V_{\underline{\ell}}\, , \qquad \cE_{\rm light} = \{ \underline{\ell}\in \cE  \, | \ \ell_1, \ \ldots , \ell_{n-1} \leq 3, \, \ell_n < 3 \}\ ,
\eeq
where we recall that $V_{\underline{\ell}}$ are the vector spaces introduced in \eqref{split-Vell}.
If we require $\mathbf{Q} \in V_{\rm light}$ we thus have a corresponding state that becomes massless 
at the infinite distance point. Let us remarks in order here. Firstly, the requirement 
$\mathbf{Q} $ is a sufficient condition for masslessness, since $\| \mathbf{Q} \|$ gives an upper bound for $M(\mathbf{Q})$. 
In other words, there could be states that become massless in the infinite distance limit whose charges 
are not in $V_{\rm light}$.\footnote{In particular, all Type F states introduced in \eqref{eq:TypeGF} become massless at the infinite distance point. Note, however, that this condition together with the BPS condition is likely very strong.} Secondly, the set $V_{\rm light}$ labels states that are massless along \textit{any} path 
in the considered growth sector. Along special paths there could be more states becoming light than captured by $V_{\rm light}$. 
Let us already mention that this path-independence requirement will be equally relevant when studying the 
 axion decay constants and thus will be discussed further in section \ref{WGCaxions}. 

Next we want to make a distinction between two types of components for $\mathbf{Q}$, such that we can write it as 
\beq
    \mathbf{Q} =  \mathbf{Q}^{\rm G} + \mathbf{Q}^{\rm F} \ ,
\eeq
where the subscripts indicate that we are dealing with Type G states and Type F states, respectively.\footnote{In \cite{Grimm:2018ohb} Type G and F states were dubbed Type I and II states respectively. It is convenient to change the name to avoid confusion with the types of limits in \eqref{all_limits}.}   We define this split, such that the 
former are intimately linked to the presence of gravity, while the latter are also present in purely field theoretic settings. 
Concretely, we require that Type G states become massless as a power law in the $y^i$, 
whereas Type F states do so at an exponential rate.
These growth rates can be inferred from the product of the charge vectors $\mathbf{Q}$ with the nilpotent orbit $\mathbf{\Pi}_{\rm nil}$. Namely, by using the nilpotent orbit approximation for $\mathbf{\Pi}$ we neglect exactly the exponential terms that determine the mass of Type F states. Then the conditions for Type F and Type G states can be phrased as
\begin{equation}\label{eq:TypeGF}
\langle \mathbf{\Pi}_{\rm nil} , \mathbf{Q}^{\rm F} \rangle = 0 \, , \qquad \langle \mathbf{\Pi}_{\rm nil}, \mathbf{Q}^{\rm G} \rangle \neq  0 \, .
\end{equation}
Note that this does not define a unique split. The set of Type F states defines a vector space. However, $ \mathbf{Q}^{\rm G}$ should be viewed 
as representing an equivalence class, since we are free to add any Type F charges to a Type G state. 
This last feature will be crucial in the context of the axion Weak Gravity Conjecture later on. 

Although Type G states do not form a vector space, we can still write down a basis of representatives $\{\mathbf{\hat q}_{\hat I} \}_{\hat I = 0,1,...}$ for the Type G charges. 
However, relating such a basis to the subspaces $V_{\underline{\ell}}$ turns out more difficult. Let us therefore derive a sufficient condition 
that determines a set of Type G vectors. The main issue is that the nilpotent orbit approximation relies on the vector $\mathbf{a}_0$, whereas it is $\R\, \mathbf{\tilde{a}}_0$ and $\I \, \mathbf{\tilde{a}}_0$ that have a definite location \eqref{location_RIa0} in the spaces $V_{\underline{\ell}}$. To relate the nilpotent orbit and the Sl(2)-orbit \eqref{Sl2-orbit} directly, we must take a special limit. Namely, if we take the limit $\frac{y^1}{y^2}, \, \ldots ,\, \frac{y^{n-1}}{y^n}, \, y^n \to \infty$ we can replace $\mathbf{\Pi}_{\rm nil}$ by $\mathbf{\Pi}_{\rm Sl(2)}$. Now let us apply this to the defining property of a Type F state. Since it should hold for all values of $y^i$, we find the following implication
\begin{equation}
\langle \mathbf{\Pi}_{\rm nil} , \mathbf{Q}^F \rangle = 0 \qquad \implies \qquad \langle \mathbf{\Pi}_{\rm Sl(2)} , \mathbf{Q}^F \rangle = 0\, .
\end{equation}
We can use the negation of this statement to find Type G states. Namely, it tells us that
\begin{equation} \label{orth_Sl2}
\langle \mathbf{\Pi}_{\rm Sl(2)} , \mathbf{q} \rangle \neq 0  \qquad \implies \qquad \langle \mathbf{\Pi}_{\rm nil} , \mathbf{q} \rangle \neq 0\, ,
\end{equation}
with $\mathbf{q} \in H^3(Y_3, \mathbb{R})$. Thus the search for a basis of Type G states can be partly fulfilled by finding vectors $\mathbf{q}$ that satisfy $\langle \mathbf{\Pi}_{\rm Sl(2)} , \mathbf{q} \rangle \neq 0$. And because $\mathbf{\Pi}_{\rm Sl(2)}$ can be expressed in terms of $\mathbf{\tilde{a}}_0$, we can directly relate such vectors to the vector spaces $V_{\underline{\ell}}$. We can thus use \eqref{orth_Sl2} to derive a relevant 
set of Type G states. 

With these preliminaries, we are now able to further discuss the construction of the charge vectors $\mathbf{Q}$ for the infinite tower of states. We already noted that we will restrict our considerations to charges $\mathbf{Q} \in V_{\rm light}$ 
defined in \eqref{def-Vlight}. Furthermore we pointed out that only the Type G charge of these states is relevant in the context of the SDC.
These can be determined using the condition \eqref{orth_Sl2}, and the polarization constraint 
\beq\label{eq:polarization}
 - i^{3-d_n} \langle   \mathbf{\tilde a_0} , (N^-_{1})^{d_1} (N^-_{2})^{d_2 - d_1} \ldots (N^-_{n})^{d_n - d_{n-1}} \bar {\tilde{\mathbf{a}}}_0\rangle > 0\, ,
\eeq
where $\underline{d}=(d_1,...,d_n)$ was given in \eqref{location_RIa0}. Hence, we see that by using the $\mathfrak{sl}(2,\bbC)^n$-algebra a set of candidate vectors 
generating the Type G states is given by acting sufficiently 
many times with $N^-_1,...,N_n^-$ on the vectors $\R\, \mathbf{\tilde a_0}$ and $\I\, \mathbf{\tilde a_0}$, such that the resulting 
vectors are in $V_{\rm light}$. In order to do that it is convenient to recall that 
\beq
     N_{i}^- V_{\underline \ell}\, \subseteq\, V_{\underline{\ell}'} \quad \text{with} \quad  {\underline{\ell}'} = (\ell_1,...,\ell_{i-1},\ell_{i}-2,...,\ell_n -2)\ ,
\eeq
and the location of $\R\, \mathbf{\tilde a_0}$ and $\I\, \mathbf{\tilde a_0}$ is $V_{\underline{d} +\underline{3}}$ as given in  \eqref{location_RIa0}. Let us denote by $\{\mathbf{q}_{\hat I} \}_{\hat I = 0,1,...}$ the set of charge vectors obtained by acting with $N^-_1,...,N_n^-$ on $\R\, \mathbf{\tilde a_0},\, \I\, \mathbf{\tilde a_0}$ that are located in $V_{\rm light}$.

Having constructed representatives $\mathbf{q}_{\hat I}$ of all Type G states, 
we can check their respective growths by using \eqref{general-norm-growth}. 
By definition of $V_{\rm light}$ all such states will have decreasing norm $\| \cdot \|$ when approaching the singularity. 
Let us denote by $\mathbf{q}_0$ the state with the slowest decrease. In order to do that we consider the growth of 
a vector $\mathbf{q}_I$ to be smaller or equal to the growth of $\mathbf{q}_J$, if there exists a finite constant  $\gamma$ such that
\beq
   \frac{|| \mathbf{q}_I ||}{|| \mathbf{q}_J ||}  < \gamma\ , 
\eeq
along every path approaching $t^1,...,t^n \rightarrow i \infty$ in the considered growth sector \eqref{eq:growthsector}. It is not hard to check in examples that this gives a well-defined transitive order among the $\mathbf{q}_{\hat I}$, since they are in $V_{\rm light}$ and constructed from $\R\, \mathbf{\tilde a_0},\, \I\, \mathbf{\tilde a_0}$. It is, however, important to stress that there are cases in which there is no 
unique element with the slowest decrease. We then can consider the set of elements with the slowest decrease and pick any element calling it $\mathbf{q}_{0}$.

We have now the sufficient preparation to introduce the infinite set of charge vectors that we will consider. We thus define  
\begin{equation} \label{QG-orbit}
\mathbf{Q}^{\rm G}(\mathbf{q}_0|m_I) = \mathbf{q}_0 + \sum_I m_I \mathbf{q}_I   \, ,
\end{equation}
with $m_I$ some  integer coefficients. For simplicity, we will take the $m_I$ to be non-negative in the rest of this paper. It was argued in \cite{Grimm:2018ohb} that the tower of states relevant to the SDC 
should arise from increasing the numbers $m_I$.  The intuitive argument for this statement was by considering 
stability of BPS states, i.e.~by asking if the states $\eqref{QG-orbit}$ labelled by $m_I$ can possibly 
decay. For the general expression \eqref{QG-orbit} of $\mathbf{Q}^{\rm G}(\mathbf{q}_0|m_I)$ stability is very hard to analyze.  
However, as noted in \cite{Grimm:2018ohb} the situation improves if the charges $\mathbf{Q}^{\rm G}(m_I)$ can be represented as an orbit
\beq \label{Charge-Orbit}
        \mathbf{Q}^{\rm G}(\mathbf{q}_0|m_I) = \text{exp} \Big( \sum_I  m_I N_I^-\Big)   \mathbf{q}_0 \ .
\eeq
In this case one can argue for a stability argument by using the phase shifts and we will 
discuss this in more detail in section \ref{stability}. However, as we will see below and was already pointed out in \cite{Grimm:2018ohb}, such an orbit does not exist for every type of limit. 
In these cases one can still write down a tower of states \eqref{QG-orbit} labelled by integers $m_I$ by using 
several Type G states with the same growth. However, in such cases one loses the stability arguments valid for the orbit \eqref{Charge-Orbit} 
and different arguments would have to be employed.\footnote{This was also stressed in \cite{Joshi:2019nzi}.} We 
stress here, that a slightly more involved the construction of \eqref{QG-orbit} and \eqref{Charge-Orbit} was suggested in \cite{Grimm:2018cpv}. Namely, it 
was shown that there exists a natural construction of the orbit and hence the $\mathbf{q}_I$ if the type of the singularity 
enhances further when sending more than $n$ coordinates to a limit. We will not need this construction in the following when working 
with \eqref{QG-orbit}.

Having discussed the Type G component $\mathbf{Q}^{\rm G}$ of the charge vectors, let us next turn to the Type F 
component $\mathbf{Q}^{\rm F}$. As mentioned above, we are in principle free to add any Type F charges to our charge vector, since these will only result in additional terms for the central charge $Z(\mathbf{Q})$ that are exponentially suppressed. Thus from the perspective of the SDC, we do not have to keep track of such components of the state. However, it will be crucial from the perspective of the axion WGC to include these Type F charges for the state. We can therefore consider the following generalized charge vector 
\begin{equation}\label{Q-charge}
\mathbf{Q}(\mathbf{q}_0| m_I, \, m_{\underline{\ell}})  = \mathbf{Q}^{\rm G}(\mathbf{q}_0|m_I)+\mathbf{Q}^{\rm F}(m_{\underline{\ell}}) = 
\mathbf{q}_0 + \sum_I m_I \mathbf{q}_I + \sideset{}{'}\sum_{\underline{\ell} \in \cE_{\rm light},\, \alpha_{\underline{\ell}}} m_{\underline{\ell}}^{\alpha_{\underline{\ell}}} \mathbf{v}^{\underline{\ell}}_{\alpha_{\underline{\ell}}}
\end{equation}
where the prime indicate that we should exclude the basis vectors $\mathbf{q}_0, \mathbf{q}_I$ in the sum over the Type F charge vectors. This means that the Type F charges $m_{\underline{\ell}}^{\alpha_{\underline{\ell}}}$ exclude these components as well. 

\subsection{On the stability of the D3-brane states}\label{stability}
Now that we have discussed which D3-brane states become light in the infinite distance limit, we want to examine the stability properties of this tower of states. In \cite{Grimm:2018ohb} it was already found for one-parameter Type IV infinite distance limits that, at a given instance along the limit, only a finite number of these states are stable against decays. Let us denote this finite number by $m_{\rm crit}^I$ for the charge generated by $\mathbf{q}_I$, which indicates the critical length of our tower of states. They found that this critical length scaled as $m_{\rm crit}^1 \sim y^1$, such that the length of the tower increases as we move further along the infinite distance limit, and that the tower becomes of infinite size as we send $y^1 \to \infty$. Here we will generalize this feature to multi-parameter infinite distance limits, which will play an important role in the test of the axion Weak Gravity Conjecture in section~\ref{WGCaxions}.

The arguments made in \cite{Grimm:2018ohb} relied crucially on aspects of $\mathcal{N}=2$ BPS states, and thus so will ours. Let us therefore begin by shortly recalling their stability properties, and refer to the original articles \cite{Joyce:1999tz,Kachru:1999vj,Douglas:2000qw,Denef:2000nb,Douglas:2000gi,Denef:2001xn,Aspinwall:2001zq,Aspinwall:2002nw,Jafferis:2008uf,Aspinwall:2009qy,Andriyash:2010yf}. Consider three BPS states $A,B$ and $C$, with their charge vectors denoted by $\mathbf{q}_A, \mathbf{q}_B$ and $\mathbf{q}_C$ respectively. We want to study the situation where state $C$ is unstable against decay into anti-state $\bar{A}$ and state $B$. The charge vectors of these states must satisfy
\begin{equation}\label{eq:chargeconservation}
\mathbf{q}_{C} = \mathbf{q}_{\bar{A}} + \mathbf{q}_B\, ,
\end{equation}
with $\mathbf{q}_{\bar{A}} = - \mathbf{q}_{A}$. The masses of these states satisfy
\begin{equation}
M(\mathbf{q}_C) \leq M(\mathbf{q}_{\bar{A}}) + M(\mathbf{q}_B)\, ,
\end{equation}
which follows from the linearity of the central charge in the charges. The state $C$ then becomes unstable against decay if this inequality is satisfied. This statement can be made more explicit by considering the (normalized) phase of the central charge
\begin{equation}
\varphi = \frac{1}{\pi} \Im \log Z(\mathbf{q})\, .
\end{equation}
The alignment of the phases $\varphi(B)=\varphi(\bar{A})$ then indicates that $C$ becomes unstable against the decay into $\bar{A}$ and $B$, which corresponds to
\begin{equation}
    \varphi(B)-\varphi(A) =1\, .
\end{equation}
The loci of this equation in $\cM^{\rm cs}$ are called curves of marginal stability, such that if we cross this curve in the moduli space, the state $C$ is only marginally stable against decay. In essence, this boils down to a restriction of this phase to a range $(-1 ,1)$ in order for the state $C$ to remain stable, as discussed in more detail in \cite{Douglas:2000qw, Denef:2001xn} in the context of so-called stable pairs. This aspect played an important role in the argument presented in \cite{Grimm:2018ohb}, where they generated the states of the tower by circling the infinite distance point until the phase of the central charge began to rotate and they thus crossed such a curve of marginal stability. Then the number of windings before they encountered this curve set the critical length $m_{\rm crit}^1$ of the tower.

It is important to mention that the product states $\bar{A},B$ cannot be chosen arbitrarily such that \eqref{eq:chargeconservation} holds, but they must be mutually non-local as well, that is, $\langle \mathbf{q}_{\bar{A}}, \mathbf{q}_B \rangle \neq 0$ \cite{Joyce:1999tz,Kachru:1999vj}. Phrased oppositely, it tells us that two states can only form a bound state if their charges are mutually non-local. Then for states that lie in $V_{\rm light}$ it is useful to recall that, by use of orthogonality relations $\eqref{V-orth}$, this vector space satisfies the following property 
\begin{equation}
    \langle V_{\rm light}, V_{\rm light} \rangle = 0\, .
\end{equation}
Therefore no two states that lie in $V_{\rm light}$ can form a bound state together. In particular, this tells us that any two states $\mathbf{Q}(\mathbf{q}_0 | m_I, m_{\underline{\ell}} ), \mathbf{Q}(\mathbf{q}_0 | m'_I, m'_{\underline{\ell}} )$ in our infinite tower are mutually local, such that they cannot bound together. It also indicates that the product states that result from the decay of a state $\mathbf{Q}(\mathbf{q}_0 | m_I, m_{\underline{\ell}} )$ of our tower must necessarily have charges that do not lie in $V_{\rm light}$ to have a non-zero intersection product between these states.

Before we argue for stability properties of our tower of states, it will prove to be useful to examine the asymptotic properties of the central charge of Type G charge vectors in more detail. To be more precise, we want to consider Type G charges that are constructed out of $\mathbf{ \tilde a}_0$ by applying lowering operators $N_i^-$, which will always be realized in our constructions. 
Therefore, we want to rewrite the nilpotent orbit \eqref{Pinil} using the $\mathfrak{sl}(2,\bbC)$-data. This can be done by rewriting the orbit as \cite{CKS, AST_1989__179-180__67_0}
\begin{equation} \label{Pnil_SL}
\mathbf{\Pi}_{\rm nil} =  \alpha(x,y) \, e^{-x^i N_i} e^{-1}(y)   \,    p(y) \, e^{-i N_{(n)}^-} \mathbf{\tilde{a}}_0\, ,
\end{equation}
where $\alpha(x,y)$ is some overall coefficient function that accounts for the freedom to rescale the periods.  
Crucially, the nilpotent orbit contains the matrix-valued function
\bea
e(y) &=& \exp\Big[  \frac{1}{2} \sum_{j=1}^{n-1} \log\Big( \frac{y^{j}}{y^{j+1}} \Big) Y_{(j)} + \frac{1}{2} \log(y^n)Y_{(n)}  \Big] \\
 &=& \Big( \frac{y^1}{y^2}\Big)^{\frac{1}{2} Y_{(1)}} ... \Big(\frac{y^{n-1}}{y^n} \Big)^{\frac{1}{2} Y_{(n-1)}}
 (y^n)^{\frac{1}{2} Y_{(n)}} \ , \nn
\eea
which encodes the asymptotic behaviour in the limit $y^1,...,y^n \rightarrow \infty$ and is the origin of the scaling in \eqref{general-norm-growth}. The complex matrix-valued function $p(y)$ is a polynomial in $(y^1/y^2)^{-1/2}$, $...$, $(y^{n-1}/y^{n})^{-1/2}$, $(y^n)^{-1/2}$ with constant term $1$. Note that the other terms in this polynomial can be bounded by factors of $\lambda^{-1/2}$ within the growth sector \eqref{eq:growthsector}.

Then consider the central charge $Z(\mathbf{q})$ of a Type G charge $\mathbf{q}$ that is located in a single eigenspace $V_{\underline{\ell}}$. For simplicity let us set $x=0$, since we want to determine the growth in the coordinates $y^i$. Then we obtain
\begin{equation}
    Z(\mathbf{q}) = \alpha(x,y) \, e^{K/2} \langle e^{-1}(y)     \,  p(y) \, e^{-i N_{(n)}^-} \mathbf{\tilde{a}}_0, \mathbf{q} \rangle \, .
\end{equation}
By properties of the symplectic product $\langle \cdot, \cdot \rangle$ we can move $e^{-1}$ to the other side of the product as $e$, and application on $\mathbf{q}$ then results in the same growth behavior as one finds for  $\| \mathbf{q}\|$ using \eqref{general-norm-growth}. Inserting \eqref{Pnil_SL} into the expression for $e^{K/2}$ we then obtain
\begin{equation}\label{Zasymptotic}
    \boxed{\quad \rule[-.3cm]{0cm}{.9cm} Z(\mathbf{q} ) \sim e^{i\theta}  \Big( \frac{y^1}{y^2}\Big)^{\frac{\ell_{1}-3}{2}} ... \Big(\frac{y^{n-1}}{y^n} \Big)^{\frac{\ell_{n-1}-3}{2}}
 (y^n)^{\frac{\ell_{n}-3}{2}} \langle p(y) e^{-i N_{(n)}^-} \mathbf{ \tilde a}_0, \mathbf{q} \rangle \, , \quad} 
\end{equation}
where $\theta$ is the overall phase inherited from $\alpha$ and we stress that this equation is true only up to overall numerical factors. 
Note that for relative phases between two central charges the factor $e^{i\theta}$ cancels, such that the only remaining part of the phase is determined by $\langle p(y) e^{-iN_{(n)}^-} \mathbf{\tilde a}_0, \mathbf{q} \rangle $.  

To simplify this expression even further, consider a Type G charge vector $\mathbf{q}$ that is constructed out of $\mathbf{\tilde a}_0$ by acting with lowering operators $N_i^-$. We can infer from the polarization condition \eqref{eq:polarization} that the intersection product of $e^{-i N_{(n)}^-} \mathbf{\tilde{a}}_0$ via the constant $1$ in the expansion of $p(y)$ with $\mathbf{q}$ is non-zero by construction.\footnote{Other terms in this expansion of $p(y)$ can potentially lead to non-zero intersections as well, but these contributions can never exceed the growth of $Z(\mathbf{q})$ that follows from the constant term 1, since these other terms are bounded by factors of $\lambda^{-1/2}$.} Then \eqref{Zasymptotic} reduces to 
\begin{equation}
 Z(\mathbf{q} ) \sim e^{i\theta}  \Big( \frac{y^1}{y^2}\Big)^{\frac{\ell_{1}-3}{2}} ... \Big(\frac{y^{n-1}}{y^n} \Big)^{\frac{\ell_{n-1}-3}{2}}
 (y^n)^{\frac{\ell_{n}-3}{2}} \langle  e^{-i N_{(n)}^-} \mathbf{ \tilde a}_0, \mathbf{q} \rangle \, .
\end{equation}
In this case $\langle e^{-iN_{(n)}^-} \mathbf{\tilde a}_0, \mathbf{q} \rangle $ indicates the relevant part of the phase of the central charge.

Now we want to investigate how far we can move up into our tower before states start to become unstable, where we will set the Type F charges to zero for simplicity. Thus we consider a state  $\mathbf{q}_0$, and look at how the phase of its central charge $\varphi(\mathbf{q}_0)$ shifts as we move up in the tower to a non-zero value $m_J  $ for a Type G charge. Then we find that
\begin{equation}\label{eq:phaseshift}
\begin{aligned}
| \varphi \big(\mathbf{Q}^{\rm G}(m_J ) \big) - \varphi(\mathbf{q}_0 ) | &=  \frac{1}{\pi}\left\lvert \Im \log\Big[ 1+\frac{Z(m_J \mathbf{q}_J)}{Z(\mathbf{q}_0)}     \Big] \right\rvert \\
&\sim  \frac{m_J \|\mathbf{q}_J \|}{\pi \| \mathbf{q}_0 \|} \left\lvert \Im \frac{\langle e^{-iN_{(n)}^-} \mathbf{\tilde a}_0, \mathbf{q}_J \rangle }{ \langle e^{-iN_{(n)}^-} \mathbf{\tilde a}_0, \mathbf{q}_0 \rangle} \right\rvert\\
& \lesssim \frac{m_J \|\mathbf{q}_J \|}{\pi \| \mathbf{q}_0 \|}
\end{aligned}
\end{equation}
where we expanded this logarithm and used \eqref{Zasymptotic}. Note the analogy with \cite{Grimm:2018ohb}, where they circled the infinite distance loci to generate the tower of states, which resulted in these phase shifts. This expression is simply the generalized version of those phase shifts to multi-parameter infinite distance limits, which reduces to the one-parameter result $\| \mathbf{q}_0\| / \| \mathbf{q}_1 \| \sim y^1$ by picking $\mathbf{q}_0 = (N_1^-)^2 \mathbf{\tilde a}_0$ and $\mathbf{q}_1 = (N_1^-)^3 \mathbf{\tilde a}_0$ using \eqref{general-norm-growth}. It hints at some critical scale 
\begin{equation}\label{mcrit}
    m_{\rm crit}^J \sim \| \mathbf{q}_0 \| / \| \mathbf{q}_J \|\, ,
\end{equation}
such that the phase of the central charge $Z(\mathbf{Q}^{\rm G}(\mathbf{q}_0|m_J)$ potentially shifts by $\frac{1}{2}$. If this phase shift actually occurs, depends on whether the central charges of $\mathbf{q}_0$ and $\mathbf{q}_I$ differ in phase, as can be seen in the second line of \eqref{eq:phaseshift}. For charge vectors given in orbit form \eqref{Charge-Orbit} this phase difference is always realized, since $\mathbf{q}_J = N_J^- \mathbf{q}_0$, and thus the numerator picks up a factor of $i$ less than the denominator by use of the polarization condition \eqref{eq:polarization}. For charge vectors that cannot be written in orbit form, one needs to go more carefully through the polarization conditions \eqref{eq:polarization}.

We can then see this shift in phase of the central charge $Z(\mathbf{Q}^{\rm G}(m_J))$ as we increase $m_J$ as an indication that the states will become unstable after a certain critical scale $m_{\rm crit}^J$. Namely, we know from the stability properties of BPS states that a state can become unstable when phases of central charges rotate, and we have
\begin{equation}
\varphi(\mathbf{Q}^{\rm G}(m_J)) = \begin{cases} \varphi(\mathbf{q}_0)+\frac{1}{2}\, , &\text{ for $\| \mathbf{q}_0 \|/\| \mathbf{q}_J \| \ll m_J$,} \\
\varphi(\mathbf{q}_0)\, , &\text{ for $\| \mathbf{q}_0 \|/\| \mathbf{q}_J \| \gg m_J$.} 
\end{cases}
\end{equation}
Thus the phase of $Z(\mathbf{Q}^{\rm G}(m_J))$ must rotate in the region given by $\| \mathbf{q}_0 \| / \| \mathbf{q}_J \| \sim m_J$, such that relative phase between product states can potentially rotate out of the stable range $(-1,1)$, and $\mathbf{Q}^{\rm G}(m_I)$ then becomes unstable. Therefore it is a sufficient condition to require $m_J \lesssim m_{\rm crit}^J \sim \| \mathbf{q}_0 \| / \| \mathbf{q}_J \|$ to ensure the stability of the states that we consider at a given instance along the limit. Note that different arguments would have to be employed for the growth of the tower in the case that the norms of $\mathbf{q}_0$ and $\mathbf{q}_J$ have the same growth behavior. However, for the purposes of this work such properties will not be needed.

We expect a similar story to hold for the Type F charges of our states, that is, the charges $m_{\underline{\ell}}$ must have some upper bound $m_{\rm crit}$ as well. This bound can be motivated from the fact that their contribution to the central charge vanishes asymptotically by construction \eqref{eq:TypeGF}, since Type F charges enter the central charge via the exponentially suppressed corrections to the nilpotent orbit approximation \eqref{Pinil}. Then the upper bound $m_{\rm crit}$ for Type F charges should grow such that contributions to the central charge still vanish asymptotically, which suggests scales comparable to exponential growth.

\subsection{One-parameter infinite distance limits}

It is instructive to briefly review the properties of the towers of states that arise in one-parameter infinite distance limit. This means that we consider Type II$_b$, III$_c$, and IV$_d$ infinite distance limits with one parameter $y^1$. A first task is to determine the dimensions of the 
vector spaces $V_\ell$. Using appendix \ref{appVsplit1d} and the classification in \cite{Kerr2017,Grimm:2018cpv}, we readily find the result listed in 
Table \ref{type-split}.

\newcolumntype{V}{>{\centering\arraybackslash} m{.21\linewidth} }
\newcolumntype{L}{>{\arraybackslash} m{.21\linewidth} }
\begin{table}[h!]
\centering
\begin{tabular}{| c  | ccc | c |}
\hline
name & $\dim V_0$ & $\dim V_1$ & $\dim V_2$ & $\mathbf{q}_0, \mathbf{q}_1$ \\
\hline \hline
\rule[-0.6cm]{0cm}{1.4cm} $\mathrm{I}_a$ &
   0&0& $a$
  & 
  \begin{minipage}{.2\textwidth} -  \end{minipage}
\\ \hline
\rule[-0.5cm]{0cm}{1.2cm}$\mathrm{II}_b$  & 0 & 0 & $b+2$ 
  & 
  \begin{minipage}{.2\textwidth} $\mathbf{q}_0 = N_1^- \text{Re}( \mathbf{\tilde{a}}_0) $  \\ $\mathbf{q}_1 = N_1^- \text{Im}( \mathbf{\tilde{a}}_0) $  \end{minipage}
 \\ \hline
\rule[-0.5cm]{0cm}{1.2cm} $\mathrm{III}_c$ & 0 & $2$ & $c$ 
  & 
  \begin{minipage}{.2\textwidth} $\mathbf{q}_0 = (N_1^-)^2 \text{Re}( \mathbf{\tilde{a}}_0)$  \\ $\mathbf{q}_1 = ( N_1^-)^2 \text{Im}( \mathbf{\tilde{a}}_0)$   \end{minipage}
\\ \hline
\rule[-0.5cm]{0cm}{1.2cm}$\mathrm{IV}_d$   & $1$ & $0$ & $d$ 
  & 
  \begin{minipage}{.2\textwidth} $\mathbf{q}_0 =( N_1^-)^2 \mathbf{\tilde{a}}_0$  \\ $\mathbf{q}_1= ( N_1^-)^3 \mathbf{\tilde{a}}_0$   \end{minipage}
  \\
  \hline
\end{tabular}
\caption{This table lists the dimensions of the spaces $V_{\ell}$ for $\ell \in \cE_m = \{0,1,2 \}$ for the possible one-parameter infinite distance limits. In addition, it includes the vectors that span the Type G charges for the charge vector $\mathbf{Q}^G$. }\label{type-split}
\end{table}

The next step is to find the vectors that generate the Type G charges. As discussed in the section \ref{sec:constructionofstates} they satisfy 
\begin{equation}
    \langle \mathbf{\Pi}_{\rm Sl(2)}, \mathbf{q} \rangle \neq 0 \, .
\end{equation}
Here $\mathbf{\Pi}_{\rm Sl(2)}$ can be expressed in terms of nilpotent matrices $N_1^-$ and the vector $\mathbf{\tilde{a}}_0$. Therefore it will prove to be useful to recall some polarization conditions for this vector $\mathbf{\tilde{a}}_0$. To be begin with, we know that
\begin{equation}
    -i^{3-d} \langle \tilde{\mathbf{a}}_0, (N_1^-)^{d} \bar {\tilde{\mathbf{a}}}_0 \rangle > 0 \, ,
\end{equation}
with $d=1,2,3$ for a Type II, III or IV infinite distance limit respectively. Furthermore, we have that $\bar{\tilde{\mathbf{a}}}_0 =  \tilde{\mathbf{a}}_0$
for a Type IV infinite distance limit, such that $\tilde{\mathbf{a}}_0$ is real. And for a Type II or III singularity we have an additional polarization condition that tells us that
\begin{equation}
    \langle \tilde{\mathbf{a}}_0, (N_1^-)^{d} \tilde{\mathbf{a}}_0 \rangle = 0 \, .
\end{equation}
Together these relations suffice to identify basis vectors $\mathbf{q}_0, \mathbf{q}_1$ for the Type G charges of our charge vector $\mathbf{Q}$, which have been included in table \ref{type-split}. We should note that $\mathbf{q}_0, \mathbf{q}_1 \in V_{3-d}$ for Type II and III infinite distance limits, whereas $\mathbf{q}_0 \in V_{2}$ and $\mathbf{q}_1 \in V_0$ for a Type IV singularity.

Now let us discuss the stability of the tower of states for each of these infinite distance limits, that is, we will comment on the critical size $m_{\rm crit}^1$ of the tower of states at a given instance along the infinite distance limit. For Type IV infinite distance limits it has already been argued that $m_{\rm crit}^1 \sim y^1$ in \cite{Grimm:2018ohb}, which is in agreement with our discussion in section \ref{stability}. In contrast, for Type II and III infinite distance limits we cannot 
apply directly the stability arguments after \eqref{eq:phaseshift}, since $\mathbf{q}_0, \mathbf{q}_1$ have the same growth. In specific examples, one might be able to argue for an $m_{\rm crit}$ using global properties of the moduli space \cite{Corvilain:2018lgw,Joshi:2019nzi}. Fortunately, we will not need an expression for $m_{\rm crit}$ in these cases.

\section{The Weak Gravity Conjecture for R-R axions}\label{WGCaxions}

In this section we study the couplings of R-R axions in the context of the Weak Gravity Conjecture (WGC) for axions. Let us therefore shortly recall this conjecture \cite{ArkaniHamed:2006dz}. In the case that one has only one axion, the conjecture states 
 that there exists an instanton coupled to the axion such that \footnote{We should note that we have set $M_{ p}=1$ throughout the rest of this section.}
\begin{equation}\label{AWGC}
    f S \leq q M_{p} \, ,
\end{equation}
where $f$ denotes the axion decay constant, $S$ is the instanton action, and $q$ is the instanton charge. For a given axion $\xi$, with a periodic field range  $\xi \cong \xi + 2\pi$, we define its decay constant $f$ via its kinetic term
\begin{equation}\label{actionAWGC}
    \cL_{\rm kin} = - f^2 \, \partial_\mu \xi\partial^\mu \xi\, .
\end{equation}
The instanton charge is defined by noting that axion couples to the instanton via the exponential $e^{-S+i q \xi}$. This 
implies that the periodicity of this contribution is $2\pi /q$ and the field range of the canonically normalized axion $2 \pi f /q$.

The formulation of the axion WGC becomes significantly more subtle in a higher-dimensional axion space. 
In order to treat such cases requires us to introduce the vectors  
\beq\label{vectors_convexhull}
    z^\cI_a  = \sum_{\cJ} \frac{f^{\cI \cJ} q_{a \cJ} }{ S_a} \ ,
\eeq 
where $f^{\cI \cJ}$ are the inverse axion decay constants,  $S_a$ instanton action of an instanton labelled by $a$, 
and $q_{a \cJ}$ encodes the axion coupling such that the instantons are weighted by a factor $\text{exp}(-S_a +  i q_{a \cJ} \xi^\cJ)$. The axion decay constants are defined 
to diagonalize the metric of the axions as $G_{\cI \cJ} = (f^T\!\cdot\! f)_{\cI \cJ}$.
It was suggested in \cite{Cheung:2014vva,Rudelius:2015xta} that the generalization of \eqref{AWGC} is the statement that the convex hull that is spanned by $\pm \mathbf{z}_a = \pm z_a^\cI \gamma_\cI$ contains the unit ball (i.e.~with radius $M^{-1}_p$), where $\gamma_\cI$ are the basis vectors in axion space normalized such that
the axion $\xi^\cI$ has a $2\pi$ periodicity.  It is important to notice that there are various 
refinements of this conjecture which propose stronger conditions \cite{ArkaniHamed:2006dz,Heidenreich:2015nta,Heidenreich:2016aqi,Grimm:2018ohb,Andriolo:2018lvp}. Most relevant for 
us will be the statement of the strong axion WGC, which states that the convex hull condition should be 
satisfied for the $\mathbf{z}_a$ constructed from the instantons with the smallest actions. In other words, 
one orders the instantons coupling to some axion direction by the value of their action and 
only retains the one with the smallest value $S_a$ to construct the vectors \eqref{vectors_convexhull}.

Before we proceed with the analysis of the general multi-axion setup, let us already give a qualitative outline of what we can expect. In the four-dimensional Type IIA Calabi-Yau compactifications that we will be considering, the quantities $f$ and $S$ vary non-trivially over the complex structure moduli space $\cM^{\rm cs}(Y_3)$. If we take an infinite distance limit in $\cM^{\rm cs}$, we will find that $f$ and $S$ have certain growth rates in the coordinates $y^i$ by use of growth properties \eqref{general-norm-growth}. The axion WGC \eqref{AWGC} then suggests that the parametrical growth of the axion decay constant $f$ should be cancelled by the decrease of the instanton action $S$. Therefore, our task is to find, for an axion direction
with a parametrically growing decay constant, an instanton that couples to this axion with an instanton action that decreases at a sufficient rate. It is at this stage that the three-cycles of $Y_3$ discussed in the previous section become relevant. Namely, instead of wrapping D3-branes on these three-cycles to find states that become massless in infinite distance limits, these three-cycles can now host Euclidean D2-branes with a parametrically decreasing action. We will find that we can couple every R-R axion to one of these instantons, provided that its axion decay constant $f$ grows in a path-independent manner. The fact that our tower of instantons also grows at a certain rate specified by $m_{\rm crit}$, discussed in section~\ref{stability}, implies that the instanton charge $q$ grows parametrically as well, which will play a crucial role in this test of the axion WGC. Namely, we will find that the decrease of the instanton action $S$ will not always suffice to cancel the growth of the axion decay constant $f$ completely, and that then the leftover growth is matched by the growth of the instanton charge $q$.

\subsection{Asymptomatic axion decay constants for R-R axions}
For completeness, we will first recall some basic aspects of R-R axions. As mentioned before, these axions $\xi^{\cI}$ follow from expanding the R-R field $C_3$ into a basis of harmonic 3-forms $\gamma_{\mathcal{I}} $ of $Y_3$ via
\begin{equation}
C_3 = \xi^{\mathcal{I}} \gamma_{\mathcal{I}}\, ,
\end{equation}
with the kinetic terms for these fields given by
\beq \label{axion-metric2}
  \cL_{\rm kin} = G_{\cI \cJ} \ \partial_\mu \xi^\cI \partial^\mu \xi^\cJ\, , \qquad G_{\cI \cJ} = \frac{1}{2} e^{2D} \langle \gamma_\cI \, ,  * \gamma_\cJ  \rangle \, ,
\eeq
where $D$ denotes the four-dimensional dilaton, and $\langle \cdot, \cdot \rangle$ is defined in \eqref{def-<>}. Then the field range of these axions is $\xi^{\mathcal{I}} \cong \xi^{\mathcal{I}} + 2\pi$, such that the metric $G_{\cI \cJ}$ defines the axion decay constants. A suitable basis for this metric adapted to the infinite distance limit is the special threeform basis discussed in section \ref{special_basis}, since it decomposes the Hodge norm in blocks that have the same growth rate \eqref{general-norm-growth}. Let us therefore split up these fields into axions $\xi^{\underline{\ell}}_{\alpha_{\underline{\ell}}}$ corresponding to the basis vectors $\mathbf{v}^{\alpha_{\underline{\ell}}}_{\underline{\ell}}$, $\alpha_{\underline{\ell}} = 1,...,\text{dim} V_{\underline{\ell}}$, that span the vector spaces $V_{\underline{\ell}}$. Collecting all $\mathbf{v}^{\alpha_{\underline{\ell}}}_{\underline{\ell}}$ we have, by using \eqref{split-Vell}, 
a basis of $H^{3}(Y_3,\bbR)$ and the  $\xi^{\underline{\ell}}_{\alpha_{\underline{\ell}}}$ parameterize all axion directions. 
The kinetic terms for these fields are then given by
\begin{equation} \label{Lkin_real}
    \cL_{\rm kin} = \frac{1}{2}e^{2D} \sum_{\underline{\ell} , \underline{r} \in \cE} \sum_{\alpha_{\underline{\ell}},\beta_{\underline{r}}} \langle \mathbf{v}_{\underline{\ell}}^{\alpha_{\underline{\ell}}}, * \mathbf{v}_{\underline{r}}^{\beta_{\underline{r}}} \rangle \ \partial_{\mu} \xi^{\underline{\ell}}_{\alpha_{\underline{\ell}}} \, \partial^{\mu} \xi^{\underline{r}}_{\beta_{\underline{r}}} \, ,
\end{equation}
which, by using \eqref{V*V} and \eqref{general-norm-growth}, we can rewrite in the infinite distance limit into
\begin{equation} \label{Lkin_asym}
    \cL_{\rm kin} \sim \frac{1}{2}e^{2D} \sum_{\underline{\ell}  \in \cE} \sum_{\alpha_{\underline{\ell}},\beta_{\underline{\ell}}}  \Big( \frac{y^1}{y^2}\Big)^{\ell_{1}-3} ... \Big(\frac{y^{n-1}}{y^n} \Big)^{\ell_{n-1}-3}
 (y^n)^{\ell_{n}-3} \langle \mathbf{v}_{\underline{\ell}}^{\alpha_{\underline{\ell}}}, *_\infty \mathbf{v}_{\underline{\ell}}^{\beta_{\underline{\ell}}} \rangle \ \partial_{\mu} \xi^{\underline{\ell}}_{\alpha_{\underline{\ell}}}  \, \partial^{\mu} \xi^{\underline{\ell}}_{\beta_{\underline{\ell}}}  \, .
\end{equation}

Some comments about \eqref{Lkin_asym} are in order here. Firstly, 
note that this expression is an asymptotic expression for the kinetic terms, as indicated by the symbol $\sim$, 
which can be used to bound the actual field space metric. As it is equally true for \eqref{general-norm-growth}, it does not capture 
the numerical factors and in fact it does not follow from our considerations how large the numerical constants 
in \eqref{Lkin_asym} need to be chosen such that it provides  a good approximation to \eqref{Lkin_real}. In fact, the 
more precise statement it is the actual Hodge norm is bounded by the norm asymptotic norm \eqref{general-norm-growth} when multiplied by 
some finite constant depending only on how close one is to the limiting point. Accordingly our results will always only gives bounds 
with undetermined numerical coefficients. 
Secondly, we note that we have used the orthogonality \eqref{V*V} among the $V_{\underline \ell}$. This simplifies the 
result significantly, but needs to be read, with our first remark in mind, as the statement that the off-diagonal terms among different
 $V_{\underline \ell}$ are sub-dominant when considering sufficiently large $y^1,...,y^n$. Nevertheless, it divides the axions into 
 various sets that `decouple' at least when considering the dominant growth. It is clear from \eqref{Lkin_asym} that we 
 can make no such decoupling statements when considering axions coming from the same $V_{\underline \ell}$. All axions 
 from a fixed $V_{\underline \ell}$ grow with the same rate in  $y^1,...,y^n$.

Let us now look at the growth of the various terms in \eqref{Lkin_asym} in more detail. Clearly, depending on the values 
of $\underline \ell$ and the considered path in the $y^1,...,y^n$ the kinetic terms either go to zero, stay constant, 
or grow to become infinitely large in the limit. Clearly, in order to test the axion WGC \eqref{AWGC} we are interested 
in increasing kinetic terms and axion decay constants.  
In this work, however, we will restrict our considerations to axions whose decay constants grows in a certain path-independent way. More precisely, 
we will demand that $f^2$ grows for any path that resides in a growth sector such as \eqref{eq:growthsector}. This immediately implies constraints on the integers $\ell_k$. Namely, we find by inspecting 
\eqref{Lkin_asym} together with \eqref{eq:growthsector} that the basis vectors $\mathbf{v}_{\underline{\ell}}$ of the considered axions 
must be elements of
\beq \label{def-Vheavy}
  V_{\rm heavy} =  \bigoplus_{\underline{\ell} \in \cE_{\rm heavy}} V_{\underline{\ell}}\, , \qquad \cE_{\rm heavy} = \{ \underline{\ell}\in \cE  \, | \ \ell_1, \ \ldots , \ell_{n-1} \geq 3, \, \ell_n > 3 \}\ .
\eeq
Note, in particular, that $V_{\rm heavy}$ is defined in an opposite fashion compared to the vector space $V_{\rm light}$ in \eqref{def-Vlight}, which contained all vector spaces whose Hodge norm decreased along every path with \eqref{t-limit} in the growth sector. In fact, we can use these two definitions 
to decompose the vector space $H^{3}(Y_3,\bbR)$ as
\beq \label{split-H3}
    H^{3}(Y_3,\bbR) = V_{\rm light} \oplus V_{\rm heavy} \oplus V_{\rm rest}\ .
\eeq
The last part $V_{\rm rest}$ are the remaining directions in the vector space decomposition. They parametrize axions whose  
decay constants grow, decrease, or stay constant depending on the considered path approaching the limit $y^1,...,y^n \rightarrow \infty$. 
Recall 
that the products between vectors out of two spaces $V_{\underline{\ell}},V_{\underline{\ell}'}$ can only be non-zero if $\underline{\ell}+\underline{\ell}' = \underline{6}$. This implies 
\beq
   \langle V_{\rm heavy}, V_{\rm heavy} \rangle =0 \ , \qquad \langle V_{\rm light}, V_{\rm light} \rangle =0\ .
\eeq
Furthermore, we find that the vector spaces $V_{\rm heavy}$ and $V_{\rm light}$ are dual to each other under the product  
$\langle \cdot,\cdot \rangle$ and indices in the index sets $\cE_{\rm light}$ and $\cE_{\rm heavy}$ can be canonically identified. In other words, we find that for every vector $\mathbf{v}_{\rm heavy} \in V_{\rm heavy}$ there exists a vector $\mathbf{v}_{\rm light} \in V_{\rm light}$ such that
\begin{equation}\label{canonduality}
\langle \mathbf{v}_{\rm heavy}, \mathbf{v}_{\rm light} \rangle \neq 0 
\end{equation}
This canonical duality between $V_{\rm heavy}$ and $V_{\rm light}$ will be crucial in arguing that we can couple every axion direction $\mathbf{v}_{\underline{\ell}} \in V_{\rm heavy}$ to a D2-brane instanton $\mathbf{Q} \in V_{\rm light}$.

Now that the asymptotic behavior of the kinetic terms has been discussed in detail, we are ready to analyze the asymptomatic axion decay constants. From \eqref{Lkin_asym} we can deduce that only the couplings between axions in the same subspace $V_{\underline{\ell}}$ are relevant in the infinite distance limit, and that axions that reside in different subspaces decouple. Therefore the asymptomatic axion decay constants are given by
\begin{equation}\label{decay_constant}
(f_{\cI \cJ})\sim \text{diag}\Big[  \, e^{D} \Big(\frac{y^1}{y_2} \Big)^{\frac{\ell_1-3}{2}} \cdots  \Big(\frac{y^{n-1}}{y^n} \Big)^{\frac{\ell_{n-1}-3}{2}}(y^n)^{\frac{\ell_n-3}{2}} \, \hat f_{\alpha_{\underline{\ell}}, \beta_{\underline{\ell}}} \Big] \, ,
\end{equation}
where we stress that we find a matrix with blocks along the diagonal labelled by $\underline \ell$ and each index $\cI$ is 
associated with one pair $\underline {\ell},\alpha_{\underline \ell}$. 
The axion decay constants for the individual blocks are proportional to $\hat f_{\alpha_{\underline{\ell}}, \beta_{\underline{\ell}}}$ 
defined as
\begin{equation}
(\hat f^T\! \cdot\! \hat f)_{\alpha_{\underline{\ell}}, \beta_{\underline{\ell}}} =  \frac{1}{2} \langle \mathbf{v}_{\underline{\ell}}^{\alpha_{\underline{\ell}}}, *_\infty \mathbf{v}_{\underline{\ell}}^{\beta_{\underline{\ell}}} \rangle \, .
\end{equation}
As discussed already in the context of the kinetic terms and in section \ref{asymp_Hodge_norm}, this matrix remains finite in the infinite distance limit, and, in fact, does not depend on the coordinates $y^i$. Thus the asymptomatic behavior in $y^i$ of the axion decay constants is captured by the power law in \eqref{decay_constant}.

\subsection{D2-brane instantons and the axion WGC}

Having discussed the kinetic terms \eqref{Lkin_real},\eqref{Lkin_asym} and the resulting axion decay constants \eqref{decay_constant}, we now study D2-brane instantons. Recall that the world-volume action of an Euclidean D2-brane wrapping a three-cycle specified by $\mathbf{Q}$ is given by
\begin{equation}\label{D2-brane}
S_{\rm D2} = e^{-D} | Z(\mathbf{Q})|+ i \langle C_3,\mathbf{Q} \rangle \, ,
\end{equation}
with $Z(\mathbf{Q})$ being the central charge defined in \eqref{central_charge}. 
The coupling functions, such as the moduli space metric, of the effective theory thus receive corrections of the form $e^{-S_{\rm D2}}$. Expanding $C_3$ in $\xi^{\underline{\ell}}_{\alpha_{\underline{\ell}}} \mathbf{v}^{\alpha_{\underline{\ell}}}_{\underline{\ell}}$ then tells us that the charges of this instanton are given by
\begin{equation}\label{instanton_charge}
q^{\underline \ell}_{\alpha_{\underline{\ell}}}(\mathbf{Q}) =  \langle \mathbf{v}^{\alpha_{\underline{\ell}}}_{\underline{\ell}}, \mathbf{Q} \rangle\, .
\end{equation}
Let us next focus on identifying candidate D2-brane instantons which have an instanton action $S$ that decreases along the infinite distance limit. The instanton action $S$ of a Euclidean D2-brane is given by the real part of $S_{\rm D2}$ in \eqref{D2-brane}, from which we obtain
\begin{equation} \label{inst_actionR}
S = e^{-D} | Z(\mathbf{Q})| \, .
\end{equation}
The idea is to consider the charges $\mathbf{Q}$ introduced in \eqref{Q-charge} that described asymptotically massless D3-brane states in the SDC consideration of section \ref{sec:SDC}. Wrapping Euclidean D2-branes on these three-cycles then provides us with candidate instantons that can couple to the R-R axions. More specifically, the mass of these D3-brane states was previously given by $|Z(\mathbf{Q})|$, and it now gives us the instanton action $S$. Thus the fact that the D3-brane states became massless in the infinite distance limit ensures us that $S$ is decreasing as well, and we have 
\begin{equation}\label{Sbound}
S = e^{-D} | Z(\mathbf{Q})|  \sim  e^{-D} \|\mathbf{q}_0 \| \, .
\end{equation}
Note that in order to get this asymptotic expression for the central charge we have used \eqref{Zasymptotic}.
Moreover, since $\mathbf{q}_0$ is the slowest decreasing charge in $\mathbf{Q}$ the leading growth of \eqref{Zasymptotic} agrees with the growth of $\|\mathbf{q}_0 \|$ up to a finite prefactor. 

We are now ready to determine the vectors $\mathbf{z}_a$ defined in \eqref{vectors_convexhull}. 
Since our instantons are labeled by the charge vector $\mathbf{Q}$ we thus need to determine $\mathbf{z}(\mathbf{Q})$.   
Inserting \eqref{decay_constant}, \eqref{instanton_charge} and \eqref{inst_actionR} into the vectors \eqref{vectors_convexhull} we find
\begin{equation}
z^{\cI} (\mathbf{Q}) \sim  \Big(\frac{y^1}{y_2} \Big)^{\frac{3-\ell_1}{2}} \cdots  \Big(\frac{y^{n-1}}{y^n} \Big)^{\frac{3-\ell_{n-1}}{2}}(y^n)^{\frac{3-\ell_n}{2}} \sum_{\beta_{\underline{\ell}}} \frac{\hat f^{\alpha_{\underline{\ell}}, \beta_{\underline{\ell}}} q^{\underline \ell}_{\beta_{\underline{\ell}}} (\mathbf{Q})}{  | Z(\mathbf{Q})|} \ ,
\end{equation}
where as before $\cI$ is split into $\underline{\ell},\alpha_{\underline \ell}$.

In the following we will simplify our discussion by no longer indicating the block structure and hence suppress the indices 
$\alpha_{\underline \ell}, \beta_{\underline \ell}$ and the finite asymptotic 
axion decay constants $\hat f_{\alpha_{\underline{\ell}}, \beta_{\underline \ell}}$. This can be done since the asymptotic 
behavior is entirely captured by the $y^i$-dependent growth factors and we will keep this relevant information for evaluating the 
axion WGC constraint. More precisely, we will consider the vectors 
\beq \label{zsimp}
z^{\cI} (\mathbf{Q}) \sim  \Big(\frac{y^1}{y_2} \Big)^{\frac{3-\ell_1}{2}} \cdots  \Big(\frac{y^{n-1}}{y^n} \Big)^{\frac{3-\ell_{n-1}}{2}}(y^n)^{\frac{3-\ell_n}{2}} \frac{q^{\underline \ell} (\mathbf{Q})}{  | Z(\mathbf{Q})|} \ ,
\eeq
when discussing the axion WGC introduced at the beginning of this section.
Since we are only concerned with parametric control, we thus ask, if the $z^{\cI} (\mathbf{Q})$ 
is bounded from below in the directions with parametrically growing axion decay constants. 
Recall that we will be only discussing directions where this happens path-independently in a
growth sector. This implies that we consider 
\beq \label{zinheavy}
    \mathbf{z}(\mathbf{Q})  =z^{\cI} (\mathbf{Q}) \gamma_\cI \ \in \ V_{\rm heavy}\ . 
\eeq
To make the growth of \eqref{zsimp} fully explicit we next use \eqref{Sbound}
and $\mathbf{q}_0 \in V_{\underline{r}}$ in \eqref{zsimp} to determine 
\beq \label{zsimp2}
    z^{\cI} (\mathbf{Q}) \sim  \underbrace{\Big(\frac{y^1}{y_2} \Big)^{\frac{6-\ell_1-r_1}{2}} \cdots  \Big(\frac{y^{n-1}}{y^n} \Big)^{\frac{6-\ell_{n-1}-r_{n-1}}{2}}(y^n)^{\frac{6-\ell_n-r_n}{2}}}_{\textstyle \hat{z}^{\underline{\ell}}(\mathbf{q}_0)} q^{\underline \ell} (\mathbf{Q})\, ,
\eeq
where we have evaluated the growth of $\mathbf{q}_0$ using \eqref{general-norm-growth}. Therefore, in order to show that the axion WGC is not parametrically violated, it will often suffice to argue that the $y^i$-dependent pre-factor $\hat{z}^{\underline{\ell}}(\mathbf{q}_0)$ in \eqref{zsimp2} is bounded from below. 
However, in order to capture all cases it turns out to be important to also consider the growth of the tower encoded by $q^{\underline \ell} (\mathbf{Q})$.
In fact, in the next subsections we will argue that one 
can use \eqref{canonduality} together with the growth of the tower discussed in section~\ref{stability} to always identify appropriate combinations of Type G and Type F charges specifying $\mathbf{Q}$ given in \eqref{Q-charge} such 
that the convex hull condition for the multi-axion WGC is never parametrically violated.

\subsection{Axion WGC in one-parameter infinite distance limits}\label{sec:AWGCone-parameter}

To illustrate how to construct the D2-brane instantons relevant for the axion WGC, let us first look at an example before moving to more general settings. Concretely, we first consider a one-parameter Type IV$_d$ infinite distance limit. As detailed in appendix \ref{appVsplit1d} we can decompose 
$V_{\rm light}$, $V_{\rm heavy}$ and $V_{\rm rest}$ into the vector spaces $V_{\ell}$ as listed in table \ref{H3-dec0}.
\begin{table}[h!]
\centering
\begin{tabular}{| c  | c | c |} \hline
\rule[-0.2cm]{0cm}{.6cm} \hspace*{.5cm} space \hspace*{.5cm} &\hspace*{.5cm} decomposition \hspace*{.5cm} &\hspace*{.5cm} dimensions \hspace*{.5cm} \\
\hline \hline
\rule[-0.2cm]{0cm}{.6cm} $V_{\rm light}$ &  $\textcolor{red}{V_{0}} \oplus \textcolor{blue}{V_{2}}$  & $\textcolor{red}{1}+ \textcolor{blue}{d} $ \\ 
\rule[-0.2cm]{0cm}{.6cm}  $V_{\rm heavy}$ &  $\textcolor{red}{V_{6}} \oplus \textcolor{blue}{V_{4}}$  & $\textcolor{red}{1}+ \textcolor{blue}{d} $ \\
\rule[-0.2cm]{0cm}{.6cm}  $V_{\rm rest}$ &  $V_{3}$  & $2(h^{2,1}-d)$ \\
\hline
\end{tabular}
\caption{Decomposition of $H^{3}(Y_3,\bbR)$ for IV$_d$. Note that we have used different colors to indicate the dimensions of the individual subspaces.}\label{H3-dec0}
\end{table}

We can reformulate the charge vectors found in \cite{Grimm:2018ohb} as
\begin{equation}\label{QtypeIV}
\mathbf{Q}^{\rm G}(m_1) = \mathbf{q}_0 + m_1 \mathbf{q}_1 \, , \qquad  \mathbf{Q}^{\rm F}(m_{2}^{i})= \sum_{\mathbf{v}_{2}^{ i} \neq \mathbf{q}_0} m_{2}^{i} \mathbf{v}_{2}^{i} \, ,
\end{equation}
with the representatives for the Type G charges given by
\begin{equation}
\mathbf{q}_0 = (N_1^-)^2 \mathbf{\tilde{a}}_0 \in V_2\, , \qquad \mathbf{q}_1 = (N_1^-)^3 \mathbf{\tilde{a}}_0 \in V_0 \, .
\end{equation}
Note that the sum over Type F charges for $\mathbf{Q}^{\rm F}$ indeed excludes the two basis vectors $\mathbf{q}_0, \mathbf{q}_1$, and only sums over the remaining $d-1$ basis vectors for $V_2$. In the following we will discuss how these instantons $\mathbf{Q}$ allow us to 
examine the WGC for axion directions in each of the subspaces of $V_{\rm heavy}$ via the vectors $\mathbf{z}(\mathbf{Q})$.

First consider an axion direction $\mathbf{v}_{4} \in V_4$ for $\mathbf{z}(\mathbf{Q})$. Its growth in this component $z^4(\mathbf{Q})$ can be deduced from \eqref{zsimp2}, and using that $\mathbf{q}_0 \in V_2$ we find that all factors of $y^1$ cancel each other, such that $\hat{z}^{4}(\mathbf{q}_0)$ remains finite along the infinite distance limit. Then to provide evidence for the axion WGC for this direction, the only remaining thing to show is that we can ensure that one of our D2-brane instantons has non-zero charge with respect to this axion, that is, $q^4(\mathbf{Q}) = \langle \mathbf{Q}, \mathbf{v}_{4} \rangle \neq 0$. This axion direction can couple to either the Type G charge vector $\mathbf{q}_0$ or some Type F charge vector $\mathbf{v}_{2}^{i}$ of $V_2$ under the product $\langle \cdot, \cdot \rangle$, which can be seen from its orthogonality properties \eqref{V-orth}. If it couples to $\mathbf{q}_0$ then every instanton in our tower \eqref{QtypeIV} suffices to argue that the axion WGC is not violated parametrically in this direction, whereas if it couples to some $\mathbf{v}_{2}^{i}$ we must pick a non-zero Type F charge $m_{2}^{i} \neq 0$ to ensure a non-zero charge with respect to this axion.

Now consider the direction $\mathbf{v}_6 \in V_6 $. The orthogonality properties \eqref{V-orth} of $\langle \cdot, \cdot \rangle$ tell us that this axion direction couples to $\mathbf{q}_1$, since $V_6$ and $V_0$ are dual to each other and both vector spaces are one-dimensional. This axion therefore couples to one of the D2-brane instantons \eqref{QtypeIV} provided that $m_1 \neq 0$. However, at first sight this axion direction seems to violate the WGC for axions, because we find by use of \eqref{zsimp2} that the growth of $z^6(\mathbf{Q})$ is given by $\hat{z}^6(\mathbf{q}_0) =(y^1)^{-1}$, since $\mathbf{q}_0 \in V_2$. This suggests that the convex hull cannot envelop a ball of finite size in this direction, and that instead the convex hull seems to shrink along this direction as we move further along the infinite distance limit. We can resolve this issue by using that we have a tower of instantons. Namely, if we consider a D2-brane instanton with $m_1 = m_{\rm crit}^1 $, with $m^1_{\rm crit} \sim y^1$ the growth of our tower discussed in section \ref{stability}, we find that its charge grows as
\begin{equation}
q^6(\mathbf{Q}) = \langle \mathbf{Q}^{\rm G}(m_{\rm crit}^1), \mathbf{v}_6 \rangle = \langle  \mathbf{q}_0 + m_{\rm crit}^1 \mathbf{q}_1 \, , \mathbf{v}_6 \rangle = m_{\rm crit}^1  \langle \mathbf{q}_1 \, , \mathbf{v}_6 \rangle \sim y^1 \, .
\end{equation}
Then combined with the growth specified by \eqref{zsimp2}, we find that
\begin{equation}
z^6(\mathbf{Q}) \sim \hat{z}^6(\mathbf{q}_0) \, q^6(\mathbf{Q}) \sim (y^1)^{-1}  (y^{1}) \sim 1
\end{equation} 
which thus provides us with a vector $\mathbf{z}(\mathbf{Q})$ whose component in the direction $\mathbf{v}_6$ is bounded in size from below, what suffices to argue that the axion WGC is not violated parametrically in this direction either.

In summary we have thus found, for every direction in $V_{\rm heavy}$, a vector $\mathbf{z}(\mathbf{Q})$ by picking an appropriate instanton $\mathbf{Q}$ out of \eqref{QtypeIV}, such that its component in that direction is bounded in size from below along this infinite distance limit. Therefore we found a set of vectors $\mathbf{z}(\mathbf{Q})$ that span a convex hull which will always contain a certain ball of finite size. From a physics perspective, this means that we have found that the axion WGC cannot be violated parametrically by considering axion decay constants that grow path-independently in this example, since we have showed that there always exists an appropriate instanton with decreasing action that couples to such axions. The fact that our tower has a critical size $m_{\rm crit}$ which increases parametrically was crucial for arguing that the  WGC for axions cannot be violated parametrically. We will motivate this feature more generally in the next subsection, where it falls under case (2), and a demonstrative figure is also provided in Figure \ref{convex-hull}.

From the discussion in \ref{classify_limits} we know that the remaining one-parameter infinite distance limits are Type II and III limits, which turn out to be slightly less interesting than the Type IV infinite distance limit. Namely, the Type G charge vectors $\mathbf{q}_0, \mathbf{q}_1$ that generate the states are located in the same eigenspace $V_1$ for a Type III limit, or $V_2$ for a Type II limit, as can be inferred from table \ref{type-split}. Furthermore, the Hodge norm on these eigenspaces $V_1$ and $V_2$ possesses the largest decrease out of all eigenspaces for the corresponding limits by use of the growth properties \eqref{general-norm-growth}, since these eigenspaces have the lowest index.\footnote{For Type III the non-empty eigenspaces are $V_1,V_2,V_3,V_4,V_5,$ and for Type II we have $V_2,V_3,V_4$.} And since the decrease of the instanton action is determined by $\|\mathbf{q}_0 \|$ via \eqref{Sbound}, we know that the growth of any axion decay constant will be matched or even exceeded by the decrease of the instanton action, using the duality between $V_{\rm heavy}$, where the axion directions reside, and $V_{\rm light}$, where the instanton charges reside. This tells us that none of the axion directions can violate the axion WGC parametrically in these examples either. Note in particular that we do not need the parametrical growth of $m_{\rm crit}$ to fullfill the WGC for axion directions that couple to $\mathbf{q}_1$, because $\mathbf{q}_1$ lies in the same eigenspace as $\mathbf{q}_0$, whereas it played a crucial role for the Type IV infinite distance limit.

\subsection{Strategy for general infinite distance limits}
Here we argue that the axion WGC cannot be violated parametrically by the R-R axions under consideration for general infinite distance limits. In doing so, we will only need to use the reformulated expression for the charge vector \eqref{Q-charge} of the D2-brane instantons, together with some requirements on the growth rate of the size of the tower $m_{\rm crit}$ that we argued for in section \ref{stability}. We can strategically analyze the axion directions $\mathbf{v}_{\underline{\ell}} \in V_{\rm heavy}$ by breaking them down to the following four cases, where we have that:
\begin{itemize}
\item[(1)] $q^{\underline{\ell}}( \mathbf{Q}^{\rm G}(m_I))  \neq 0$ for some $m_I$, and $\hat{z}^{\underline{\ell}}(\mathbf{q}_0)$ is bounded from below;
\item[(2)] $q^{\underline{\ell}}( \mathbf{Q}^{\rm G}(m_I))   \neq 0$ for some $m_I$,  and $\hat{z}^{\underline{\ell}}(\mathbf{q}_0)$  unbounded from below;
\item[(3)] $q^{\underline{\ell}}( \mathbf{Q}^{\rm F}(m_{\underline{s}}))   \neq 0$ for some $m_{\underline{s}}$, and $\hat{z}^{\underline{\ell}}(\mathbf{q}_0)$  is bounded from below;
\item[(4)] $q^{\underline{\ell}}( \mathbf{Q}^{\rm F}(m_{\underline{s}}))  \neq 0$ for some $m_{\underline{s}}$,  and $\hat{z}^{\underline{\ell}}(\mathbf{q}_0)$  unbounded from below.
\end{itemize}
The purpose of this separation of cases is to investigate how every axion direction $\mathbf{v}_{\underline{\ell}} \in V_{\rm heavy}$ can be coupled to one of the D2-brane instantons $\mathbf{Q}$ specified by \eqref{Q-charge} such that, provided we pick the right charges, the growth of this component $z^{\underline{\ell}}(\mathbf{Q}(\mathbf{q}_0 | m_I,m_{\underline{s}})) $ can be bounded from below. Namely, by showing that we can pick a vector $\mathbf{z}(\mathbf{Q})$ for every direction in $V_{\rm heavy}$ such that its component in that direction is bounded in size from below, we know that the associated convex hull must contain a ball of finite size. The first condition then indicates whether this axion couples to our tower of D2-brane instantons via one of their Type G charges or one of their Type F charges. This coupling can be ensured via the canonical duality \eqref{canonduality} between $V_{\rm heavy}$ and $V_{\rm light}$, since the Type F and Type G charge vectors of our D2-brane instantons together span the whole of $V_{\rm light}$ by construction (see section \ref{sec:constructionofstates}), such that we only need to pick the right combination of instanton charges. Therefore we are left with analyzing the growth rate of the vectors $\mathbf{z}(\mathbf{Q}(\mathbf{q}_0,m_I,m_{\underline{s}})$ in that direction, which is given by $z^{\underline{\ell}}(\mathbf{Q}(\mathbf{q}_0|m_I,m_{\underline{s}})$ in \eqref{zsimp2}.  Then wether the growth of $\hat{z}^{\underline{\ell}}(\mathbf{q}_0)$ is bounded from below tells us directly if this component remains finite for all paths within the growth sector \eqref{eq:growthsector}. However, the growth of  $\hat{z}^{\underline{\ell}}(\mathbf{q}_0)$ is not necessarily bounded from below, such that there can be directions where the convex hull seems to shrink along the infinite distance limit. The fact that we still have to account for the possible parametrical growth of charge $q^{\underline \ell} (\mathbf{Q})$ in \eqref{zsimp2} will resolve this issue for cases (2) and (4). Below we examine these cases separately in more detail, such that we can argue that each of these cases does not violate the axion WGC parametrically.

\textbf{Case (1):} We have that $q^{\underline{\ell}}( \mathbf{Q}^{\rm G}(m_I))=\langle \mathbf{Q}^{\rm G}, \mathbf{v}_{\underline{\ell}} \rangle \neq 0$ for some $m_I$, so the axion direction couples to a Type G charge of our charge vector. Additionally, we know that the growth of $ \hat{z}^{\underline{\ell}}(\mathbf{q}_0) $ in \eqref{zsimp2} is bounded from below, such that the convex hull can envelop a ball of finite size in this direction, which indicates that these axions can therefore never violate the axion WGC parametrically. For example, the axion direction $N^-_1 \mathbf{\tilde{a}}_0 \in V_4$ in the one-parameter Type IV limit belonged to this case.

\textbf{Case (2):} This case seems to violate the axion WGC at first sight, since $\hat{z}^{\underline{\ell}}(\mathbf{q}_0) $ cannot be bounded from below, that is, $\hat{z}^{\underline{\ell}}(\mathbf{q}_0) \to 0$ for certain paths within the growth sector \eqref{eq:growthsector}, which suggests that the convex hull can shrink in this direction as we move along the infinite distance limit, similar to the axion direction $\mathbf{\tilde a}_0 \in V_6$ we encountered in the one-parameter Type IV in section \ref{sec:AWGCone-parameter}. However, we did not account for the charge of this instanton yet, which is given by\footnote{The fact that $\hat{z}^{\underline{\ell}} (\mathbf{q}_0) $ cannot be bounded from below indicates that $\|\mathbf{q}_{0} \| \| \mathbf{v}_{\underline{\ell}} \|$ can grow parametrically for certain paths, such that \eqref{general-norm-growth} tells us that the eigenspaces $V_{\underline{r}},V_{\underline{\ell}}$ in which they reside are not dual to each other under the product $\langle \cdot, \cdot \rangle$. Therefore we must have $\langle \mathbf{q}_0 , \mathbf{v}_{\underline{\ell}} \rangle =0$ by use of orthogonality properties \eqref{V-orth}.}
\begin{equation}
  q^{\underline{\ell}}(\mathbf{Q}^{\rm G}(m_I)) =   \langle \mathbf{Q}^{\rm G}(m_I), \mathbf{v}_{\underline{\ell}} \rangle  = \sum_I m_I \langle \mathbf{q}_I, \mathbf{v}_{\underline{\ell}} \rangle\, .
\end{equation}
We know that this axion must couple to a Type G charge, so we must have that $\langle \mathbf{q}_J, \mathbf{v}_{\underline{\ell}} \rangle \neq 0$ for some $J$, and thus $\mathbf{q}_J \in V_{\underline{6}-\underline{\ell}}$ by use of \eqref{V-orth}. Then increasing this $m_J$ results in increasing the charge $q^{\underline{\ell}}(\mathbf{Q})$, such that if we pick $m_J = m_{\rm crit}^J \sim \| \mathbf{q}_0 \| / \|\mathbf{q}_J \| $ following \eqref{mcrit}, we find
\begin{equation}
\begin{aligned}
 z^{\underline{\ell}}(\mathbf{Q}(\mathbf{q}|m_{\rm crit}^J))  &\sim \hat{z}^{\underline{\ell}}(\mathbf{q}_0) \,  m_{\rm crit}^J \\
&\sim  \Big(\frac{y^1}{y_2} \Big)^{\frac{6-\ell_1-r_1}{2}} \cdots  \Big(\frac{y^{n-1}}{y^n} \Big)^{\frac{6-\ell_{n-1}-r_{n-1}}{2}}(y^n)^{\frac{6-\ell_n-r_n}{2}} \frac{ \| \mathbf{q}_0 \| }{ \|\mathbf{q}_J \| }\\
&\sim 1\, ,
\end{aligned}
\end{equation}
where we used that the growths of the norms of $\mathbf{q}_0$ and $\mathbf{q}_J$ cancel the growth rate in front precisely, using that $\mathbf{q}_0 \in V_{\underline{r}}$ and $\mathbf{q}_J \in V_{\underline{6} - \underline{\ell}}$. Thus we found that the tower of D2-brane instantons grows at exactly the right rate to avoid parametrical violations of the WGC for these axion directions, since it allowed us to pick appropriate instanton charges such that the component of $\mathbf{z}(\mathbf{Q})$ in this direction is bounded in size from below, which indicates that the convex hull can envelop a ball of finite size in this direction. Note that this crucially relies on the growth of the instanton charge $q^{\underline{\ell}}(\mathbf{Q})$, which means that one cannot use only the smallest charge with respect to these axions. A depiction of this interplay between the growth of the tower and the convex hull condition of the axion WGC has been provided in figure \ref{convex-hull}. 

\textbf{Case (3):} By picking the right Type F charges $m_{\underline{s}}$ we can ensure that such an axion direction couples to a D2-brane instanton $\mathbf{Q}$ out of our tower \eqref{Q-charge}, and the fact that $\hat{z}^{\underline{\ell}}(\mathbf{q}_0)$ is bounded from below then indicates that the convex hull can envelop a ball of finite size in this direction, which tells us that these axions cannot violate the WGC.  As an example, in the one-parameter Type IV limit in \ref{sec:AWGCone-parameter} an axion direction $\mathbf{v}_{4} \in V_4$ that does not couple to the Type G charge vector $\mathbf{q}_0$ falls under this case.

\textbf{Case (4):} We know that $\mathbf{v}_{\underline{\ell}}$ must couple to one of the Type F charge vectors via 
\begin{equation}
q^{\underline{\ell}}(\mathbf{Q}^{\rm F}(m_{\underline{s}})) =\langle \mathbf{Q}^{\rm F}(m_{\underline{s}}), \mathbf{v}_{\underline{\ell}} \rangle = \sideset{}{'}\sum_{i}  m_{\underline{6}-\underline{\ell}}^{i} \langle  \mathbf{v}_{\underline{6}-\underline{\ell}}^{i}, \mathbf{v}_{\underline{\ell}} \rangle \, ,
\end{equation}
with the relevant Type F charge vectors given by $\mathbf{v}^i_{\underline{6}-\underline{\ell}}\in V_{\underline{6}-\underline{\ell}}\subset V_{\rm light}$. However, we cannot bound $\hat{z}^{\underline{\ell}}(\mathbf{q}_0)$ from below, such that we can have $\hat{z}^{\underline{\ell}}(\mathbf{q}_0)\to 0$ for certain paths to the infinite distance limit. It indicates that the convex hull shrinks along these directions, and to avoid this we must pick $m_{\underline{6}-\underline{\ell}}^{i} \sim  \|\mathbf{q}_0 \| \| \mathbf{v}_{\underline{\ell}} \|$, such that we do not have parametrical violations of the axion WGC, similar to case (2). In section \ref{stability} we argued that the upper bound $m_{\rm crit}$ for Type F charges increases at a sufficient rate to allow for such choices of Type F charge. We should note that this case did not occur in the one-parameter examples, but can occur in multi-parameter limits.

\subsection{Axion WGC in two-parameter infinite distance limits}\label{sec:AWGCtwo-parameter}
In this section we provide some examples to demonstrate the strategy outlined above. We go through all two-parameter infinite distance limits 
$t^1,t^2 \rightarrow i \infty$ and construct the relevant charge vectors. It should be stressed that in contrast to \cite{Grimm:2018cpv} we will not require 
the charge vector $\mathbf{Q}$ to be in orbit-form. This allows us to address all possible enhancements and show a general result. 

In the case of considering a limit $t^1,t^2 \rightarrow i \infty$ 
one has two log-monodromy matrices $N_1,N_2$ and hence one has two singularity types \eqref{all_limits} associated to $N_1$ and $N_{(2)} = N_1 + N_2$
according to our discussion in section \ref{classify_limits}. Denoting these types by Type A and Type B, the two-parameter configurations 
are split into all possible enhancements Type A $\rightarrow$ Type B. Going through all relevant cases we 
will determine the split \eqref{split-H3} with $V_{\rm light} $ and $V_{\rm heavy} $ defined in \eqref{def-Vlight} and \eqref{def-Vheavy}. The 
various components are given as direct sums in the vector spaces $V_{\underline \ell} \equiv V_{\ell_1\ell_2}$.
By use of appendix \ref{appVsplit2d} we first give the dimensions of the $V_{\ell_1\ell_2}$ and indicate the positions of the basis of Type G representatives $\mathbf{q}_0, \mathbf{q}_I$ in $V_{\rm light}$, such that we can write the charge vectors in the formulation introduced in section~\ref{sec:constructionofstates}. Then we consider each axion direction $v_{\ell_1 \ell_2}$ that lies in a subspace $V_{\ell_1 \ell_2}$ of $V_{\rm heavy}$ and identify the candidate D2-brane instanton that ensures that the axion WGC is not violated parametrically. In doing so, we employ the orthogonality properties \eqref{V-orth} of the product $\langle \cdot, \cdot \rangle$ and the growth properties \eqref{general-norm-growth} for the $\mathfrak{sl}(2,\mathbb{C})$-eigenspaces $V_{\ell_1 \ell_2}$. Furthermore, we point out for each of these axion directions $\mathbf{v}_{\ell_1 \ell_2}$ to which cases it belongs in the strategy outlined in the previous section. In particular, wether $\| \mathbf{v}_{\ell_1 \ell_2} \| \| \mathbf{q}_0 \|$ can grow unboundedly indicates directly, by use of \eqref{zsimp2}, if the growth of $\hat{z}^{\ell_1 \ell_2} (\mathbf{Q})$ is bounded from below, or if it is unbounded and we need to consider growth of instanton charges as well.

\subsubsection{Enhancement I$_a$ $\to$ IV$_d$}
Let us first consider the enhancement from type I$_a$ to IV$_d$, which can occur for $d=r+a$ with $r \geq 1$. 
Then $H^3(Y_3,\bbR)$ decomposes as in table \ref{H3-dec1}.
\begin{table}[h!]
\centering
\begin{tabular}{| c  | c | c |} \hline
\rule[-0.2cm]{0cm}{.6cm} \hspace*{.5cm} space \hspace*{.5cm} &\hspace*{.5cm} decomposition \hspace*{.5cm} &\hspace*{.5cm} dimensions \hspace*{.5cm} \\
\hline \hline
\rule[-0.2cm]{0cm}{.6cm} $V_{\rm light}$ &  $\textcolor{red}{V_{30}} \oplus \textcolor{blue}{V_{22}}\oplus  \textcolor{ForestGreen}{V_{32}}$  & $\textcolor{red}{1}+ \textcolor{blue}{a} + \textcolor{ForestGreen}{r}$ \\ 
\rule[-0.2cm]{0cm}{.6cm}  $V_{\rm heavy}$ &  $\textcolor{red}{V_{36}} \oplus \textcolor{blue}{V_{44}} \oplus  \textcolor{ForestGreen}{V_{34}}$  & $\textcolor{red}{1}+ \textcolor{blue}{a} + \textcolor{ForestGreen}{r}$ \\
\rule[-0.2cm]{0cm}{.6cm}  $V_{\rm rest}$ &  $V_{33}$  & $2(h^{2,1}-a-r)$ \\
\hline
\end{tabular}
\caption{Decomposition of $H^{3}(Y_3,\bbR)$ for I$_a\, \to \, $IV$_d\,$. Note again that we use colors to indicate the dimensions of individual subspaces.}\label{H3-dec1}
\end{table}

The charge vectors found in \cite{Grimm:2018cpv} can be reformulated as
\begin{equation}
\begin{aligned}
\mathbf{Q}^{\rm G}(m_1) &=  \mathbf{q}_0 + m_1 \mathbf{q}_1 \, , \\
\mathbf{Q}^{\rm F}(m_{22},  m_{32} ) &= \sum_{i} m_{22}^{i} \mathbf{v}_{22}^{i} + \sum_{\mathbf{v}_{32}^{j} \neq \mathbf{q}_0} m_{32}^{j} \mathbf{v}_{32}^{j}
\end{aligned}
\end{equation}
with the basis of representatives for the Type G charges given by
\begin{equation}
\mathbf{q}_0 = (N_2^-)^2 \mathbf{\tilde{a}}_0 \in V_{32}\, , \qquad \mathbf{q}_1 = (N_2^-)^3 \mathbf{\tilde{a}}_0\in V_{30} \, .
\end{equation}
Now let us go systematically through axion directions for each of the subspaces of $V_{\rm heavy}$:
\begin{itemize}
\item The axion direction $\mathbf{v}_{44} \in V_{44}$ must couple to some Type F charge vector $\mathbf{v}_{22}^j $. For certain paths we can have that $\| \mathbf{v}_{44} \| \| \mathbf{q}_0 \| $ can grow unboundedly, which means that we must require $m_{22}^{i}  \sim \|\mathbf{q}_0\| \| \mathbf{v}_{44} \|$ such that we do not violate the WGC for axions parametrically. These axion directions belong to case (4).
\item The axion direction $\mathbf{v}_{36} \in V_{36}$ couples to the Type G charge vector $\mathbf{q}_1$, thus we must pick $m_1 \neq 0$ such that it couples to the D2-brane instanton. Furthermore we have that $\| \mathbf{v}_{36} \| \| \mathbf{q}_0 \| $ grows unboundedly, which indicates that we must pick $m_1 = m_{\rm crit}^1 \sim \| \mathbf{q}_0 \| /  \| \mathbf{q}_{1} \| $ to avoid parametrical violations of the axion WGC. This axion direction belongs therefore to case (2).
\item The axion direction $\mathbf{v}_{34} \in V_{34}$ can couple to either the Type G charge vector $\mathbf{q}_0$ or some other Type F charge vector $\mathbf{v}_{32}^i$. For both cases we have that the growth of $\| \mathbf{q}_0 \| \| \mathbf{v}_{34} \| $ is bounded, so we cannot have a parametrical violation of the axion WGC. If it couples to $\mathbf{q}_0$ this axion belongs to case (1), whereas if it couples to some other $\mathbf{v}_{32}^i$ we must pick $m_{32}^i \neq 0$, and thus it belongs to case (3).
\end{itemize}

\subsubsection{Enhancement II$_b$ $\to$ IV$_d$}
This type of enhancement from Type II$_b$ to IV$_d$ can occur if we have $d=r+b$ with $r \geq 1$. The vector spaces $V_{\rm light}$, $V_{\rm heavy}$ and $V_{\rm rest}$ can be decomposed as in table \ref{H3-dec2}.

\begin{table}[h!]
\centering
\begin{tabular}{| c  | c | c |} \hline
\rule[-0.2cm]{0cm}{.6cm} \hspace*{.5cm} space \hspace*{.5cm} &\hspace*{.5cm} decomposition \hspace*{.5cm} &\hspace*{.5cm} dimensions \hspace*{.5cm} \\
\hline \hline
\rule[-0.2cm]{0cm}{.6cm} $V_{\rm light}$ &  $\textcolor{red}{V_{20}} \oplus \textcolor{blue}{V_{22}}\oplus \textcolor{ForestGreen}{V_{32}}$  & $\textcolor{red}{1}+ \textcolor{blue}{b} + \textcolor{ForestGreen}{r}$ \\ 
\rule[-0.2cm]{0cm}{.6cm}  $V_{\rm heavy}$ &  $\textcolor{red}{V_{46}} \oplus \textcolor{blue}{V_{44}}  \oplus \textcolor{ForestGreen}{V_{34}}$  & $\textcolor{red}{1}+ \textcolor{blue}{b} + \textcolor{ForestGreen}{r}$ \\
\rule[-0.2cm]{0cm}{.6cm}  $V_{\rm rest}$ &  $\textcolor{red}{V_{42}} \oplus\textcolor{blue}{ V_{33}} \oplus \textcolor{ForestGreen}{V_{24}}$  & $\textcolor{red}{1}+ \textcolor{blue}{2(h^{2,1}-b-r-1)} + \textcolor{ForestGreen}{1}$ \\
\hline
\end{tabular}
\caption{Decomposition of $H^{3}(Y_3,\bbR)$ for II$_b\, \to \, $IV$_d\,$. }\label{H3-dec2}
\end{table}

%

The charge vectors found in \cite{Grimm:2018cpv} can be reformulated as
\begin{equation}
\begin{aligned}
\mathbf{Q}^{\rm G}(m_1) &=  \mathbf{q}_0 + m_1 \mathbf{q}_1 \, , \\
     \mathbf{Q}^{\rm F}(m_{32}^i, m_{22}^j) &= \sum_{i} m_{32}^i \mathbf{v}_{32}^i + \sum_{\mathbf{v}_{22}^j \neq \mathbf{q}_0} m_{22}^j \mathbf{v}_{22}^j
\end{aligned}
\end{equation}
with the basis of representatives for the Type G charges given by
\begin{equation}
\mathbf{q}_0 = N_1^- N_2^- \mathbf{\tilde{a}}_0 \in V_{22}\, , \qquad \mathbf{q}_1 = N_1^- (N_2^-)^2 \mathbf{\tilde{a}}_0\in V_{20} \, .
\end{equation}
Let us go through each of the subspaces of $V_{\rm heavy}$ systematically:
\begin{itemize}
\item For $\mathbf{v}_{46} \in V_{46}$ we know that it couples to the Type G charge vector $\mathbf{q}_1$ from orthogonality conditions, since $V_{20},V_{46}$ are dual one-dimensional vector spaces. We also know that $\| \mathbf{v}_{46} \| \| \mathbf{q}_0 \| $ can grow unboundedly, and that we therefore must pick $m_1 = m_{\rm crit}^1 \sim \| \mathbf{v}_{46} \| \| \mathbf{q}_0 \| $. Thus this axion direction belongs to case (2).
\item For $\mathbf{v}_{44} \in V_{44}$ we can have that it couples either to the Type G charge vector $\mathbf{q}_0$ or to some other Type F charge vector $\mathbf{v}_{22}^j$. Either way the growth rates of $\| \mathbf{q}_0 \|$ and $\|\mathbf{v}_{44} \|$ cancel each other. In the first case we have that it couples to Type G charge and thus belongs to (1), and in the other case it couples to Type F charge and thus belongs to (3).
\item For $\mathbf{v}_{34} \in V_{34}$ we know that it couples to some Type F charge vector $\mathbf{v}_{32}^i$, and that the growth rates of $\| \mathbf{v}_{34}\|$ and $\| \mathbf{q}_0 \|$ cancel each other. Therefore these axion directions belong to case (3).
\end{itemize}

\subsubsection{Enhancement III$_c$ $\to$ IV$_d$}
The enhancement from type III$_c$ to  IV$_d$ can occur if we have $d=r+c+2$ 
and $r \geq 0$. The vector spaces can then be decomposed as listed in table \ref{H3-dec3}.

\begin{table}[h!]
\centering
\begin{tabular}{| c  | c | c |} \hline
\rule[-0.2cm]{0cm}{.6cm} \hspace*{.5cm} space \hspace*{.5cm} &\hspace*{.5cm} decomposition \hspace*{.5cm} &\hspace*{.5cm} dimensions \hspace*{.5cm} \\
\hline \hline
\rule[-0.2cm]{0cm}{.6cm} $V_{\rm light}$ &  $\textcolor{red}{V_{10}} \oplus \textcolor{blue}{V_{12}} \oplus \textcolor{ForestGreen}{V_{22}} \oplus \textcolor{magenta}{V_{32}}$  & $\textcolor{red}{1}+ \textcolor{blue}{1} + \textcolor{ForestGreen}{c}+\textcolor{magenta}{(r+1)}$ \\ 
\rule[-0.2cm]{0cm}{.6cm}  $V_{\rm heavy}$ &  $\textcolor{red}{V_{56}} \oplus \textcolor{blue}{V_{54}}  \oplus \textcolor{ForestGreen}{V_{44}} \oplus \textcolor{magenta}{V_{34}}$  & $\textcolor{red}{1}+ \textcolor{blue}{1} + \textcolor{ForestGreen}{c}+\textcolor{magenta}{(r+1)}$  \\
\rule[-0.2cm]{0cm}{.6cm}  $V_{\rm rest}$ &  $V_{33}$  & $2(h^{2,1}-c-r-2)$ \\
\hline
\end{tabular}
\caption{Decomposition of $H^{3}(Y_3,\bbR)$ for III$_c\, \to \, $IV$_d\,$. }\label{H3-dec3}
\end{table}

The charge vectors can be given by
\begin{equation}
\begin{aligned}
\mathbf{Q}^{\rm G}(m_1, m_2) &=  \mathbf{q}_0 + m_1 \mathbf{q}_1 + m_2 \mathbf{q}_2 \, , \\
\mathbf{Q}^{\rm F}(m_{32}^i, m_{22}^j) &= \sum_{\alpha_{22}} m_{22}^j \mathbf{v}_{22}^j + \sum_{\mathbf{v}_{32}^i \neq \mathbf{q}_0} m_{32}^i \mathbf{v}_{32}^i \, ,
\end{aligned}
\end{equation}
with the basis of representatives for the Type G charges given by
\begin{equation}
\mathbf{q}_0 = N_1^- N_2^- \mathbf{\tilde{a}}_0 \in V_{32}\, , \qquad \mathbf{q}_1 = (N_1^-)^2 \mathbf{\tilde{a}}_0\in V_{12} \, , \qquad \mathbf{q}_2 = N_2^- (N_1^-)^2 \mathbf{\tilde{a}}_0\in V_{10} \, .
\end{equation}
Note that these charge vectors $\mathbf{Q}^{\rm G}$ differ from the ones found in \cite{Grimm:2018cpv}. There they generated the tower by acting with $e^{N_2^-}$ on $\mathbf{q}_1$, such that only $\mathbf{q}_1,\mathbf{q}_2$ were used to span the charges of the states. A non-zero component for $\mathbf{Q}$ in the direction $\mathbf{q}_0$ will be necessary to couple certain axion directions in $V_{34}$ to the tower of D2-brane instantons.

Now let us go through each of the subspaces of $V_{\rm heavy}$:
\begin{itemize}
\item For $\mathbf{v}_{56} \in V_{56}$ we know that the axion couples to the Type G charge $\mathbf{q}_2$. We can increase $m_2 = m_{\rm crit}^2 \sim \|\mathbf{q}_0 \| \|\mathbf{v}_{56} \|$, such that we cancel the growth of $\|\mathbf{q}_0 \| \|\mathbf{v}_{56} \| $. This axion direction belongs therefore to case (2).
\item For $\mathbf{v}_{54} \in V_{54}$ we know that it couples to the Type G charge $\mathbf{q}_1$ from orthogonality conditions. Again we have that $\|\mathbf{q}_0 \| \|\mathbf{v}_{54} \| $ can grow unboundedly, thus we must increase $m_1 = m_{\rm crit}^1 \sim \|\mathbf{q}_0 \| \|\mathbf{v}_{54} \|$ to cancel this growth. Therefore this axion direction belongs to case (2) as well.
\item For $\mathbf{v}_{44} \in V_{44}$ we have that it couples to some Type F charge $\mathbf{v}_{22}^j$. The growth rate of $\|\mathbf{v}_{44}\|$ can exceed the growth of $\|\mathbf{q}_0\|$. We must therefore increase $m_{22}^j \sim \|\mathbf{v}_{44}\| \|\mathbf{q}_0\|$. This tells us that these axion directions belong to case (4).
\item For $\mathbf{v}_{34} \in V_{34}$ we can have that it either couples to the Type G charge $\mathbf{q}_0$ or to some other Type F charge $\mathbf{v}_{32}^i$. For both cases we have that $\| \mathbf{v}_{34} \| \| \mathbf{q}_0 \| $ is bounded. If it couples to $\mathbf{q}_0$ it belongs to case (1), whereas if it couples to some $\mathbf{v}_{32}^i$ we must pick $m_{32}^i \neq 0$ and it belongs to case (3).
\end{itemize}

\subsubsection{Enhancement II$_b$ $\to$ III$_c$}
The enhancement Type II$_b$ to Type III$_c$ can occur if we have $c=b+r-2$ and $r \geq 0$. 
The vector spaces are then decomposed as in table \ref{H3-dec4}.

\begin{table}[h!]
\centering
\begin{tabular}{| c  | c | c |} \hline
\rule[-0.2cm]{0cm}{.6cm} \hspace*{.5cm} space \hspace*{.5cm} &\hspace*{.5cm} decomposition \hspace*{.5cm} &\hspace*{.5cm} dimensions \hspace*{.5cm} \\
\hline \hline
\rule[-0.2cm]{0cm}{.6cm} $V_{\rm light}$ &  $\textcolor{red}{V_{21}} \oplus\textcolor{blue}{ V_{22}}  \oplus \textcolor{ForestGreen}{V_{32}}$  & $\textcolor{red}{2} +\textcolor{blue}{(b-2)} +\textcolor{ForestGreen}{r}$ \\ 
\rule[-0.2cm]{0cm}{.6cm}  $V_{\rm heavy}$ &  $\textcolor{red}{V_{45}}  \oplus \textcolor{blue}{V_{44}} \oplus \textcolor{ForestGreen}{V_{34}}$  & $\textcolor{red}{2} +\textcolor{blue}{(b-2)} +\textcolor{ForestGreen}{r}$  \\
\rule[-0.2cm]{0cm}{.6cm}  $V_{\rm rest}$ &  $\textcolor{red}{V_{43}} \oplus \textcolor{blue}{V_{33}} \oplus \textcolor{ForestGreen}{V_{23}}$  & $\textcolor{red}{2}+\textcolor{blue}{2(h^{2,1}-b-r-1)}+\textcolor{ForestGreen}{2}$ \\
\hline
\end{tabular}
\caption{Decomposition of $H^{3}(Y_3,\bbR)$ for II$_b\, \to \, $III$_c\,$. }\label{H3-dec4}
\end{table}
%
Here we consider charge vectors different from \cite{Grimm:2018cpv}, given by
\begin{equation}
\begin{aligned}
\mathbf{Q}^{\rm G}(m_1) &=  \mathbf{q}_0 +  m_1 \mathbf{q}_1  \, , \\
 \mathbf{Q}^{\rm F}(m_{32}^i, m_{22}^j) &=  \sum_{j} m_{22}^j \mathbf{v}_{22}^j + \sum_{i} m_{32}^i \mathbf{v}_{32}^i \, ,
\end{aligned}
\end{equation}
and the basis for the representatives for the Type G charges can be given by
\begin{equation}
\begin{aligned}
\mathbf{q}_0 &= N_1^- N_2^- \Re \mathbf{\tilde{a}}_0 \in V_{21}\, , \qquad    &\mathbf{q}_1  &= N_1^- N_2^-  \Im \mathbf{\tilde{a}}_0 \in V_{21}\, .
\end{aligned}
\end{equation}
Then let us go through each of the subspaces of $V_{\rm heavy}$ for the axion directions:
\begin{itemize}
    \item The axion direction $\mathbf{v}_{45} \in V_{45}$ can couple to Type G charge vectors $\mathbf{q}_0$ and $\mathbf{q}_1$. In both cases the growth of $\| \mathbf{v}_{45} \|$ is matched by the decrease of $\| \mathbf{q}_0 \|$. These axion directions therefore belong to case (1).
    \item The axion direction $\mathbf{v}_{44} \in V_{44}$ couples to some Type F charge $\mathbf{v}_{22}^j$. The growth rate of $\| \mathbf{v}_{44} \|$ can never exceed the decrease of $\| \mathbf{q}_0 \|$, and thus we can just pick $m_{22}^j \neq 0$. Therefore these axion directions belong to case (3).
    \item The axion direction $\mathbf{v}_{34} \in V_{34}$ couples to a Type F charge vector $\mathbf{v}_{32}^i$. We have that the growth of $\| \mathbf{v}_{34} \|$ never exceeds the decrease of $\| \mathbf{q}_0 \|$. Thus we pick $m_{32}^i \neq 0$ and therefore such axion directions belong to case (3). 
\end{itemize}

\section{Conclusions}

In this paper we studied the  axion Weak Gravity Conjecture for asymptotic regimes in 
field space that are at infinite geodesic distance. Specifically we focused on the axions arising from the 
R-R three-forms in Type IIA string theory compactified on a Calabi-Yau threefold. The kinetic terms 
of these axions depend non-trivially on the complex structure moduli of the threefold, but we showed 
the this dependence can be made explicit in asymptotic regimes that are at infinite geodesic distance. 
The infinite distance points in general Calabi-Yau threefold moduli spaces that are obtained by sending any number 
of coordinates to a limit can be classified \cite{Kerr2017,Grimm:2018cpv}. We have shown that the data 
characterizing a limit can also be used to group the axions into subsets, with each subset having a
kinetic term with a common growth behaviour. We then focused on the axions that have growing 
kinetic terms for any path, in a growth sector of the form \eqref{eq:growthsector}, and hence growing axion decay constants. 
In order that these do not violate the axion Weak Gravity Conjecture, instantons have to 
become relevant with actions decreasing with the inverse rate when approaching the infinite distance point. 
By using recent insights about the SDC \cite{Grimm:2018ohb, Grimm:2018cpv}, we have argued that one can always find such 
instantons, since an infinite number of candidate D2-brane states has vanishing action at the infinite 
distance point. 

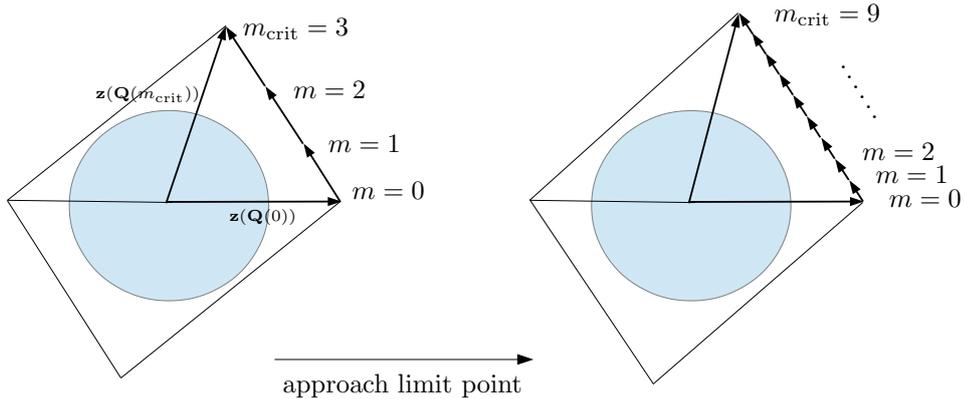
\begin{figure}[h!]
\vspace*{-.2cm}
\begin{center} 
\includegraphics[width=12cm]{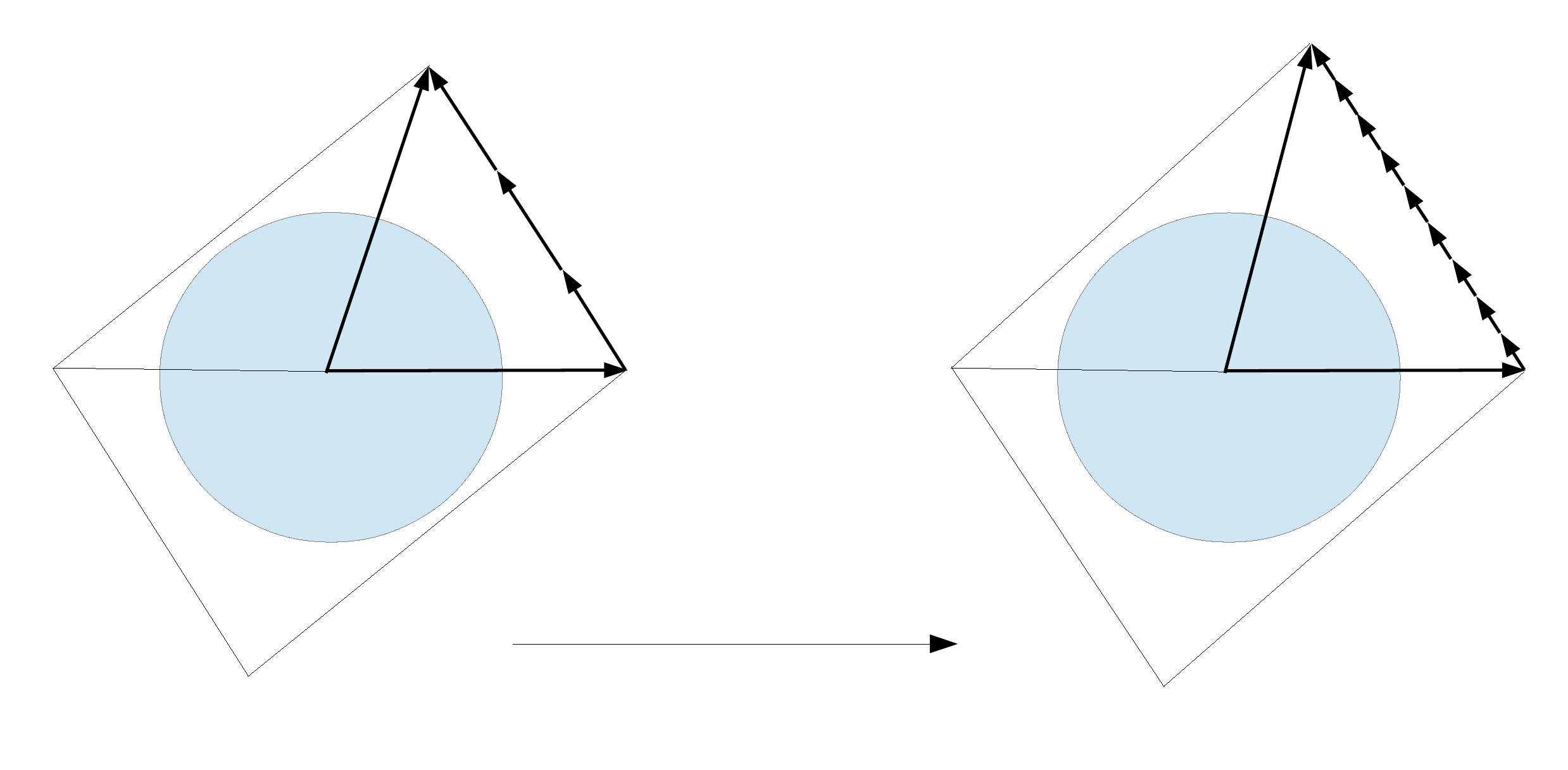} 
\vspace*{-1cm}
\end{center}
\begin{picture}(0,0)
\put(148,77){\tiny $\mathbf{z}(\mathbf{Q}(0))$}
\put(98,123){\tiny $\mathbf{z}(\mathbf{Q}(m_{\rm crit}))$}
\put(194,85){\small $m = 0$}
\put(185,103){\small $m=1$}
\put(172,123){\small $m = 2$}
\put(153,147){\small $ m_{\rm crit}=3$}
\put(395,82){\small $m=0$}
\put(390,91){\small $m = 1$}
\put(385,100){\small $m = 2$}
\put(375,135){\rotatebox{-58}{\small $\cdots \cdots $}}
\put(352,152){\small $m_{\rm crit} = 9$}
\put(168,12){\small approach limit point}
\end{picture}
\caption{Depiction of the convex hull spanned by the vectors $\pm \mathbf{z}(\mathbf{Q}(0))$ and $\pm \mathbf{z}(\mathbf{Q}(m_{\rm crit}))$, which correspond to the lowest and highest instanton in our tower respectively. Note that we explicitly depicted the steps that we go through as we move up in this tower of instantons. As we move further along the infinite distance limit, we find that these steps become smaller and smaller, which is compensated by the fact that our tower of instantons is becoming larger, such that the total length of this side of the convex hull remains finite and thus that the convex hull will always contain a ball of finite size.} \label{convex-hull}
\end{figure}

In order to address the axion WGC for multiple axions, we have constructed a set of 
vectors $\mathbf{z}(\mathbf{Q})$ that depend on the axion decay constants, the instanton action, and instanton charge. Here it was crucial to introduce appropriate 
charge vectors $\mathbf{Q}(\mathbf{q}_0 | \underline{m})$ in \eqref{Q-charge} such that these instantons 
actually correct the effective theory.
The convex hull of such 
vectors should contain the unit ball, in order that the axion WGC is satisfied. We stress that 
this statement does not have to be true for the smallest  charge coupling the instanton to the 
axion. Indeed we have shown that there are many infinite distance limits in moduli space, namely the limits 
that contain type IV enhancements, for which the convex hull of the $\mathbf{z}(\mathbf{Q})$ cannot contain 
a unit ball if one considers the smallest instanton charge. The emerging picture is, however, compelling: 
the closer we approach such infinite distance 
 points the higher the instanton charge of the instanton relevant in \eqref{vectors_convexhull} has to be. 
 We have depicted this result in figure \ref{convex-hull}. This implies that actually an ever increasing tower of instanton 
 states becomes relevant when approaching the infinite distance point. Clearly, such a picture is 
 reminiscent of the SDC where an increasing tower of particles needs to be included in the effective theory. 
 It is interesting that our findings can also be viewed as providing evidence for the strong axion WGC, if the 
 charge vectors $\mathbf{Q}(\mathbf{q}_0 | \underline{m})$ are indeed describing the 
 lightest stable states relevant for the SDC. The instanton actions for all charges 
 $\mathbf{Q}(\mathbf{q}_0 | \underline{m})$ have the same leading growth determined by $\mathbf{q}_0$ and are thus 
 equally relevant in the effective theory. It would be very interesting to explore this further and, in particular, 
 clarify the role of individual Type G and Type F states that are not of the form $\mathbf{Q}(\mathbf{q}_0 | \underline{m})$.
 
In our analysis of the D2-brane instantons it was crucial to collect information about D-brane states with asymptotically 
vanishing actions. We started our considerations by asserting that these states are BPS and their action 
can be determined by evaluating the central charge. Furthermore, at least for a certain large class of possible limits, 
it was essential to argue how the number of stable states changes when approaching the infinite distance point. In order 
to do that we generalized the stability argument of \cite{Grimm:2018ohb} to multiple variables. More precisely, we 
derived a maximal growth of the tower of states that are stable when approaching the infinite distance point
by ensuring the absence of decays of these states. Our stability arguments apply to situations in which one 
can identify a charge vector $\mathbf{q}_0$ that has a distinguished slowest decrease of the associated 
central charge. Such an identification was also important in the construction of the charge orbits of \cite{Grimm:2018cpv}
to which our results naturally apply. It is important to stress, however, that even if the instanton charges do not take the form of a charge orbit 
our findings suffice to provide general evidence that the  axion WGC is not parametrically violated. 
It should be clear that our findings cannot be conclusive when 
it comes to checking the BPS properties of D-brane states. It would be interesting to explore these issues 
further and, in particular, study stability at limits in moduli space that do not allow for a local construction 
of a charge orbits. A potential avenue was suggested in \cite{Grimm:2018cpv}, and exemplified in \cite{Corvilain:2018lgw}, in which 
charge orbits were transferred along the moduli space. 

Our results were obtained in full generality for any infinite distance limit in complex structure moduli space 
and hence do not apply to only a specific example or a class 
of examples. This was achieved by using the powerful mathematical machinery 
of \cite{Schmid,CKS}, which describes so-called limiting mixed Hodge structures. One of the 
central results of these papers is the introduction of $n$ 
commuting $\frak{sl}(2,\bbC)$-algebras associated to an infinite distance locus obtained
by sending $n$ coordinates to a limit. These algebras act on the vector space $H^{3}(Y_3,\bbC)$
and induce a canonical splitting of $H^{3}(Y_3,\bbR)$, which we argued to be crucial in studying the axion kinetic 
terms. The $\frak{sl}(2,\bbC)$-data arises from log-monodromy matrices $N_i$, the limiting period vector $\mathbf{a}_0$, 
and the growth sector \eqref{eq:growthsector} associated to the path along which one takes the limit. 
We believe that this approach will be fruitful in many further applications \cite{inprep}. 
However, it should be stressed that it is particularly powerful 
when it comes to estimates, such as the ones encountered for the axion WGC. This can be traced back 
to the fact that the asymptotic behaviour of periods $\mathbf{\Pi}$ can be bounded by using 
the $\frak{sl}(2,\bbC)$-structure, but the corrections are only under parametric control. 
It would be very interesting to systematically classify the corrections arising in the link of 
$\mathbf{\Pi}$ to the $\frak{sl}(2,\bbC)$-structure.

This leaves us to close with highlighting further interesting open problems for future projects. 
A first direction is to address the generalization of our considerations to any path in complex structure 
moduli space and hypermultiplet moduli space. On the one hand, this would require 
to go beyond the growth sector description presented here. On the other hand, it would also amount to consider 
paths in which one sends the four-dimensional 
dilaton $e^D$ to a limit. Satisfying the axion WGC then requires to consider more general D-brane configurations as  
very recently also discussed in \cite{Marchesano:2019ifh,Font:2019cxq,Lee:2019xtm}. We stress that it is an interesting and challenging task to unify the mathematical 
structure presented here, with the general insights about the hypermultiplet moduli space \cite{Alexandrov:2011va,Alexandrov:2013yva}. 
A second open question is to address the issue of emergence in hypermultiplet moduli space. 
More precisely, it was suggested in \cite{Grimm:2018ohb}, that infinite distances in moduli space could be emergent 
from integrating out the infinite tower of states relevant to the SDC. Furthermore, it was very recently 
argued in \cite{Marchesano:2019ifh} that in certain situations the inclusion of D-instanton corrections 
into the moduli space metric can render formerly infinite distance points to lie at finite distance. 
It would be interesting to check if this is indeed true for all infinite distance limits investigated here.  
Finally, let us close with the rather obvious statement that we did not check the precise numerical 
constraint suggested by the axion WGC. While the introduced mathematical machinery gives bounds 
on the relevant quantities, the appearing numerical coefficients are not further constraint. While they 
can be derived in explicit examples, it would be very exciting to check if there are general constraints arising from geometry.

\subsection*{Acknowledgements}

It is a pleasure to thank Chongchuo Li, Eran Palti,  Irene Valenzuela, and Stefan Vandoren
for valuable discussions.

\appendix

\begingroup
\allowdisplaybreaks

\section{Derivation of eigenspaces for infinite distance limits}\label{appVsplit}
In this appendix we decompose the eigenspaces $V_{\underline{\ell}}$ in the primitive subspaces $P^{p,q}(N_{(k)})$ for one- and two-parameter infinite distance limits, including their dimensions. We make the location of the vectors that follow from $\mathbf{\tilde{a}}_0 \in P^{3,d_n}(N_{(n)}^-)$ explicit,\footnote{Recall that $d_n = 0,1,2,3$ for an $n$-parameter Type I$_a$, II$_b$, III$_c$ or IV$_d$ infinite distance limit respectively.} because of their importance in constructing Type G charge vectors in sections \ref{sec:SDC} and \ref{WGCaxions}. These results can be argued from the Hodge-Deligne diamonds in each step of the enhancement chain, which were given in \cite{Grimm:2018cpv}. We explain this procedure for the one- and two-parameter infinite distance limits separately. Note that the $V_{\underline \ell}$ given in this appendix are the complexifications of the vector spaces used in the main text.

\subsection{Eigenspaces $V_{\ell}$ for one-parameter infinite distance limits}\label{appVsplit1d}
Here we give the decomposition and the dimension of the eigenspaces $V_{\ell}$ for all types of one-parameter infinite distance limits considered in section \ref{sec:SDC}. The content of these spaces can be read off from the rows of the Hodge-Deligne diamond which were given in \cite{Grimm:2018ohb, Grimm:2018cpv}, with the index of the row indicating the subscript $\ell$. This follows from the fact that elements of the same row have the same eigenvalue under the generator $Y_{1}$ of the $\mathfrak{sl}(2,\mathbb{C})$-triple.

\subsubsection{Type I$_a$}
For a Type I$_a$ infinite distance limit the eigenspaces $V_{\ell}$ and their dimensions are given by
{\small
\begin{align}
V_{4} &=  P^{2,2}(N_1^-)\, , \qquad  & \dim V_{4} &= a\, ,\nn \\[-.1cm]
V_{3} &= P^3(N_1^-)\, ,\qquad & \dim V_{3} &= 2(a'+1)\, ,\\[-.1cm]
V_2 &=  N_1^- P^{2,2}(N_1^-)\, , \qquad & \dim V_{2} &= a\, .\nn
\end{align}}
Note that $\mathbf{\tilde{a}}_0 \in P^{3,0}(N_1^-) \subseteq P^3(N_1^-)$.

\subsubsection{Type II$_b$}
For a Type II$_b$ infinite distance limit the eigenspaces $V_{\ell}$ and their dimensions are given by
{\small
\begin{align}
V_{4} &= \spanC{\mathbf{\tilde{a}}_0, \mathbf{\bar{\tilde{a}}}_0} \oplus P^{2,2}(N_1^-)\, ,\qquad  & \dim V_{4} &= b+2\, ,\nn\\[-.1cm]
V_{3} &= P^3(N_1^-)\, ,\qquad  & \dim V_{3} &= 2b' \, ,\\[-.1cm]
V_2 &= \spanC{N_1^- \mathbf{\tilde{a}}_0, N_1^- \mathbf{\bar{\tilde{a}}}_0} \oplus N_1^- P^{2,2}(N_1^-)\, ,\qquad  & \dim V_{2} &= b+2\, .\nn
\end{align}}
Note that $\mathbf{\tilde{a}}_0 \in P^{3,1}(N_1^-) \subseteq P^4(N_1^-)$.
\subsubsection{Type III$_c$}
For a Type III$_c$ infinite distance limit the eigenspaces $V_{\ell}$ and their dimensions are given by
{\small
\begin{align}
V_5 &=  \spanC{\mathbf{\tilde{a}}_0, \mathbf{\bar{\tilde{a}}}_0} \, , \qquad & \dim V_{5} &= 2\, ,\nn\\[-.1cm]
V_{4} &= P^{2,2}(N_1^-)\, , \qquad  & \dim V_{4} &= c\, ,\nn\\
V_{3} &= \spanC{N_1^- \mathbf{\tilde{a}}_0,N_1^-  \mathbf{\bar{\tilde{a}}}_0} \oplus P^3(N_1^-)\, ,\qquad  & \dim V_{3} &= 2(c'+1) \, ,\\[-.1cm]
V_2 &= N_1^- P^{2,2}(N_1^-)\, , \qquad & \dim V_{2} &= c\, ,\nn\\[-.1cm]
V_1 &=  \spanC{(N_1^-)^2 \mathbf{\tilde{a}}_0, (N_1^-)^2  \mathbf{\bar{\tilde{a}}}_0} \, , \qquad & \dim V_{1} &= 2\, .\nn
\end{align}}
Note that $\mathbf{\tilde{a}}_0 \in P^{3,2}(N_1^-) \subseteq P^5(N_1^-)$.
\subsubsection{Type IV$_d$}
For a Type IV$_d$ infinite distance limit the eigenspaces $V_{\ell}$ and their dimensions are given by
{\small
\begin{align}
V_6 &=  \spanC{\mathbf{\tilde{a}}_0} \, , \qquad  & \dim V_{6} &= 1\, ,\nn\\[-.1cm]
V_{4} &= \spanC{N_1^- \mathbf{\tilde{a}}_0} \oplus P^{2,2}(N_1^-)\, , \qquad  & \dim V_{4} &= d+1\, ,\nn\\[-.1cm]
V_{3} &= P^3(N_1^-)\, , \qquad  & \dim V_{3} &= 2d' \, ,\\
V_2 &= \spanC{(N_1^-)^2 \mathbf{\tilde{a}}_0} \oplus N_1^- P^{2,2}(N_1^-)\, , \qquad  & \dim V_{2} &= d+1\, ,\nn\\[-.1cm]
V_0 &=  \spanC{(N_1^-)^3 \mathbf{\tilde{a}}_0} \, , \qquad  & \dim V_{0} &= 1\, .\nn
\end{align}}
Note that $\mathbf{\tilde{a}}_0 \in P^{3,3}(N_1^-) \subseteq P^6(N_1^-)$.

\subsection{Eigenspaces $V_{\ell_1 \ell_2}$ for two-parameter infinite distance limits}\label{appVsplit2d}
Here we give the spaces $V_{\ell_1 \ell_2}$ for the two-parameter infinite distance limits considered in section \ref{sec:AWGCtwo-parameter}. Let us first shortly explain the procedure used to derive these spaces $V_{\ell_1 \ell_2}$, which uses the Hodge-Deligne diamonds considered in \cite{Grimm:2018cpv}. We start with the Hodge-Deligne diamond induced by $N_1^-$. This diamond can be split up into components $(N_1^-)^a P^b(N_1^-)$, with $P^b(N_1^-)$ the primitive vector space of weight $b$ associated with $N_1^-$. The row to which this component belongs determines the number $\ell_1= b-2a$, similar to the one-parameter infinite distance limits. Then $N_2^-$ induces a mixed Hodge structure on these components $(N_1^-)^a P^b(N_1^-)$ individually. In practice, this means that $(N_1^-)^a P^b(N_1^-)$ is split up into further pieces, and each piece belongs to a specific point in the Hodge-Deligne diamond induced by $N_{(2)}^-$. Then we can determine $\ell_2$ from the row at which this piece ends up in the Hodge-Deligne diamond induced by $N_{(2)}^-$. 

\subsubsection{Enhancement I$_a$ $\to$ IV$_d$}
Here we consider the enhancement of a Type I$_a$ infinite distance limit to a Type IV$_d$ infinite distance limit by sending an additional coordinate to infinity, i.e. $y^2 \to \infty$. The decompositions and dimensions of the eigenspaces $V_{\ell_1  \ell_2}$ are then given by
{\small
\begin{align}
V_{44} &=   P^{2,2}(N_1^-)\, , \qquad &\dim(V_{44}) &= a\, , \nn\\[-.1cm]
V_{36} &= \spanC{\mathbf{\tilde{a}}_0}\, , \qquad &\dim(V_{36}) &= 1\, ,\nn\\[-.1cm]
V_{34} &= \spanC{N_2^- \mathbf{\tilde{a}}_0   } \oplus \Big( P^3(N_1^-) \cap P^{2,2}(N_{(2)}^-) \Big) \, , \qquad &\dim(V_{34}) &= r\, , \nn\\[-.1cm]
V_{33} &= P^3(N_{(2)}^-) \, , \quad &\dim(V_{33}) &= 2(a'-r)\, ,\\[-.1cm]
V_{32} &=\spanC{(N_2^-)^2 \mathbf{\tilde{a}}_0   } \oplus  N_2^- \Big( P^3(N_1^-) \cap P^{2,2}(N_{(2)}^-)\Big)\, , \qquad &\dim(V_{32}) &= r\, ,\nn\\[-.1cm]
V_{30} &= \spanC{(N_2^-)^3 \mathbf{\tilde{a}}_0} \, , \qquad &\dim(V_{30}) &= 1 \, ,\nn\\[-.1cm]
V_{22} &=  N_1^- P^{2,2}(N_1^-)\, , \qquad &\dim(V_{22}) &= a \, .\nn
\end{align}}
Note that $\mathbf{\tilde{a}}_0 \in P^{3,3}(N^-_{(2)})$.

\subsubsection{Enhancement II$_b$ $\to$ IV$_d$}
Here we consider the enhancement of a Type II$_b$ infinite distance limit to a Type IV$_d$ infinite distance limit. The decompositions and dimensions of the eigenspaces $V_{\ell_1  \ell_2}$ are then given by
{\small
\begin{align}
V_{46} &= \spanC{\mathbf{\tilde{a}}_0} \, , \qquad &\dim(V_{46}) &= 1 \, ,\nn\\[-.1cm]
V_{44} &= \spanC{N_2^-\mathbf{\tilde{a}}_0} \oplus \Big( P^{2,2}(N_1^-)\cap P^{2,2}(N_{(2)}^-) \Big) \, , \qquad &\dim(V_{44}) &= b \, ,\nn\\[-.1cm]
V_{42} &= \spanC{(N_2^-)^2 \mathbf{\tilde{a}}_0}  \, , \qquad &\dim(V_{42}) &= 1\, ,\nn\\[-.1cm]
V_{34} &=  P^3(N_1^-) \cap P^{2,2}(N_{(2)}^-) \, , \qquad &\dim(V_{34}) &= r\, ,\nn\\[-.1cm]
V_{33} &= P^3(N_{(2)}^-)\, , \qquad &\dim(V_{33}) &= 2(b'-r)\, ,\\[-.1cm]
V_{32} &= N_2^- \Big(P^3(N_1^-) \cap P^{2,2}(N_{(2)}^-)\Big) \, , \qquad &\dim(V_{32}) &= r\, ,\nn\\[-.1cm]
V_{24} &= \spanC{N_1^- \mathbf{\tilde{a}}_0} \, , \qquad &\dim(V_{24}) &=1 \, ,\nn\\[-.1cm]
V_{22} &= \spanC{N_1^- N_2^-\mathbf{\tilde{a}}_0} \oplus  N_1^- N_2^- \Big( P^{2,2}(N_1^-)\cap P^{2,2}(N_{(2)}^-) \Big)\, , \ \ &\dim(V_{22}) &= b\, ,\nn\\[-.1cm]
V_{20} &= \spanC{(N_1^- (N_2^-)^2 \mathbf{\tilde{a}}_0}\, , \qquad &\dim(V_{33}) &= 1\, .\nn
\end{align}}
Note that $\mathbf{\tilde{a}}_0 \in P^{3,3}(N^-_{(2)})$.

\subsubsection{Enhancement III$_c$ $\to$ IV$_d$}
Here we consider the enhancement of a Type III$_c$ infinite distance limit to a Type IV$_d$ infinite distance limit. The decompositions and dimensions of the eigenspaces $V_{\ell_1  \ell_2}$ are then given by
{\small
\begin{align}
V_{56} &= \spanC{\mathbf{\tilde{a}}_0}\, , \qquad &\dim(V_{56}) &= 1\, ,\nn\\[-.1cm]
V_{54} &= \spanC{N_2^{-} \mathbf{\tilde{a}}_0}\, ,\qquad &\dim(V_{54}) &= 1\, ,\nn\\[-.1cm]
V_{44} &= P^{2,2}(N_1^-) \, ,\qquad &\dim(V_{44}) &= c\, ,\nn\\[-.1cm]
V_{34} &= \Big( P^{3}(N_1^-)\cap P^{2,2}(N_{(2)}^-) \Big) \oplus \spanC{N_1^- \mathbf{\tilde{a}}_0}\, ,\qquad &\dim(V_{34}) &= r+1\, ,\nn\\[-.1cm]
V_{33} &= P^3(N_{(2)}^-)\, ,\qquad &\dim(V_{33}) &= 2(c'-r-1)\, ,\\[-.1cm]
V_{32} &=  N_2^- \Big( P^{3}(N_1^-)\cap P^{2,2}(N_{(2)}^-) \Big) \oplus \spanC{N_1^- N_2^- \mathbf{\tilde{a}}_0}\, ,\quad &\dim(V_{32}) &= r+1\, ,\nn\\[-.1cm]
V_{22} &= N_1^- P^{2,2}(N_1^-)\, ,\qquad &\dim(V_{33}) &= c\, ,\nn\\[-.1cm]
V_{12} &= \spanC{(N_1^-)^2\mathbf{\tilde{a}}_0}\, , \qquad &\dim(V_{12}) &= 1\, ,\nn\\[-.1cm]
V_{10} &= \spanC{N_2^-(N_1^-)^2\mathbf{\tilde{a}}_0} \qquad &\dim(V_{10}) &= 1\, .\nn
\end{align}}
Note that $\mathbf{\tilde{a}}_0 \in P^{3,3}(N^-_{(2)})$.

\subsubsection{Enhancement II$_b$ $\to$ III$_c$}
Here we consider the enhancement of a Type II$_b$ infinite distance limit to a Type III$_c$ infinite distance limit. The decompositions and dimensions of the eigenspaces $V_{\ell_1  \ell_2}$ are then given by
{\small
\begin{align}
V_{45} &= \spanC{\mathbf{\tilde{a}}_0, \mathbf{\bar{\tilde{a}}}_0} \, ,  \qquad &\dim(V_{45}) &= 2\, ,\nn\\[-.1cm]
V_{44} &= P^{2,2}(N_1^-) \, , \qquad &\dim(V_{44}) &= b-2\, ,\nn\\[-.1cm]
V_{43} &= \spanC{N_2^- \mathbf{\tilde{a}}_0, N_2^- \mathbf{\bar{\tilde{a}}}_0} \, , \qquad &\dim(V_{43}) &= 2\, ,\nn\\[-.1cm]
V_{34} &= P^3(N_1^-) \cap P^{2,2}(N_{(2)}^-) \, , \qquad &\dim(V_{34}) &=r \, ,\nn\\[.1cm]
V_{33} &= P^3(N_{(2)}^-) \, , \qquad &\dim(V_{33}) &= 2(b'-r)\, ,\\[-.1cm]
V_{32} &= N_2^- \Big(  P^3(N_1^-) \cap P^{2,2}(N_{(2)}^-) \Big) \, , \qquad &\dim(V_{32}) &= r\, ,\nn\\[.1cm]
V_{23} &= \spanC{N_1^- \mathbf{\tilde{a}}_0, N_1^- \mathbf{\bar{\tilde{a}}}_0} \, , \qquad &\dim(V_{23}) &= 2\, ,\nn\\[-.1cm]
V_{22} &= N_1^- P^{2,2}(N_1^-) \, , \qquad &\dim(V_{22}) &= b-2\, ,\nn\\[-.1cm]
V_{21} &= \spanC{N_1^- N_2^- \mathbf{\tilde{a}}_0, N_1^- N_2^- \mathbf{\bar{\tilde{a}}}_0} \, , \qquad &\dim(V_{33}) &= 2\, .\nn
\end{align}}
Note that $\mathbf{\tilde{a}}_0 \in P^{3,2}(N^-_{(2)})$.
\endgroup

\bibliographystyle{JHEP}
\bibliography{bibliography}

\end{document}